\documentclass[aip,pop,reprint,amsmath,amssymb]{revtex4-1}

\usepackage{graphicx}
\usepackage{floatrow}

\usepackage{bm}
\usepackage{xfrac}
\usepackage{color}
\usepackage{mathtools}
\usepackage{array,multirow}
\usepackage{float}
\usepackage{cancel}
\usepackage{upgreek}
\usepackage{siunitx}
\usepackage{esvect}
\usepackage{xcolor}

\definecolor{themeRed}{RGB}{146,0,0}
\definecolor{themeGreen}{RGB}{0,146,146}
\definecolor{themeOrange}{RGB}{219,109,0}
\definecolor{themePurple}{RGB}{73,0,146}
\definecolor{themeGray}{RGB}{146,146,146}
\definecolor{themePink}{RGB}{255,109,182}
\definecolor{themeBlue}{RGB}{109,182,255}

\usepackage{tikzexternal}

\tikzsetfigurename{Figure}
\makeatletter
\expandafter\gdef\csname c@tikzext@no@\tikzexternal@figurename\endcsname{1}%
\def\pgfexternal@restore#1{#1}
\usepackage{stackengine}

\setstackgap{L}{1.4cm}
\stackMath

\newcommand{\myannotate}[3][2cm]{\stackunder{\mathmakebox[#1][c]{\displaystyle #3}}{\parbox[b]{#1}{\footnotesize \center#2 \vspace{.4cm}}}}

\usepackage[colorlinks,
    linkcolor={blue!80!black},
    citecolor={blue!80!black},
    urlcolor={blue!80!black}]{hyperref}
\usepackage[capitalise]{cleveref}
\setcitestyle{numbers,square}%

\newcommand\diff{\mathop{}\!\mathrm{d}}

\newcommand\etc{etc\@ifnextchar.{}{.\@}}
\newcommand\etal{et al\@ifnextchar.{}{.\@}}
\makeatother

\begin{document}


\title{Incorporating Kinetic Effects on Nernst Advection in Inertial Fusion Simulations}


\author{J.~P.~Brodrick} 
\email{jonathan.brodrick@york.ac.uk}
\affiliation{York Plasma Institute, Dep't of Physics, University of York, Heslington, York, YO10 5DD, United Kingdom}
\author{M.~Sherlock}
\author{W.~A.~Farmer}
\affiliation{Lawrence Livermore National Laboratory, 7000 East Ave., Livermore, CA, USA}
\author{A.~S~Joglekar}
\affiliation{University of California Los Angeles, CA, USA}
\author{R.~Barrois}
\affiliation{University of Technology, Eindhoven, The Netherlands}
\author{J.~Wengraf}
\affiliation{University of Manchester, UK}
\author{J.~J.~Bissell}
\affiliation{University of Bath, UK}
\author{R.~J.~Kingham}
\affiliation{Plasma  Group, Blackett Laboratory, Imperial College London, London SW7 2AZ, United Kingdom}
\author{D.~Del~Sorbo}
\author{M.~P.~Read}
\author{C.~P.~Ridgers}
\affiliation{York Plasma Institute, Dep't of Physics, University of York, Heslington, York, YO10 5DD, United Kingdom}

\date{\today}

\begin{abstract}
We present a simple method to incorporate nonlocal effects on the Nernst advection of magnetic fields down steep temperature gradients, 
and demonstrate its effectiveness in a number of inertial fusion scenarios.
This is based on assuming that the relationship between the Nernst velocity and the heat flow velocity is unaffected by nonlocality.
The validity of this assumption is confirmed over a wide range of plasma conditions by comparing Vlasov-Fokker-Planck and flux-limited classical transport simulations.
Additionally, we observe that the Righi-Leduc heat flow is more severely affected by nonlocality due to its dependence on high velocity moments of the electron distribution function, but are unable to suggest a reliable method of accounting for this in fluid simulations.
\end{abstract}

\pacs{}

\maketitle

\newcommand{\capsformat}[1]{\textsc{#1}}


\section{Introduction}

Recent advances in indirect drive laser fusion at the National Ignition Facility (\capsformat{nif}) using high-density carbon ablators and low gas-fill \cite{ablatorComparison} have now led to
neutron yields in excess of $10^{16}$ \cite{BerzakHopkins}.
However, the ignition frontier of net energy gain (with respect to the total laser energy) remains elusive.
 One particular challenge, 
is that there are significant discrepancies between experimental results and models 
\cite{Hurricane},
which have, until very recently, required ad hoc multipliers
on the radiation drive to avoid overestimating it
by up to 30\% \cite{Jones}.
This recent elimination of drive multipliers would not have been possible were it not for large reduction of another tunable parameter---the flux-limiter---which 
is used to approximate reductions in the electron heat flux from 
nonlocal effects, self-generated magnetic fields and plasma instabilities.
Our paper aims to address the crossover between the first two areas through detailed comparisons with fully-kinetic Vlasov-Fokker-Planck (\capsformat{vfp}) simulations, while also considering accurate and efficient ways to account for  
nonlocal modifications to the `Nernst' advection of magnetic fields down temperature gradients.

Developments in proton radiography \cite{NonMono,ProtonRadio} and magnetohydrodynamic 
modelling capabilities \cite{Davies,Farmer2017} have encouraged a resurgence of interest in the role of magnetic fields in inertial confinement fusion (\capsformat{icf}). Self-generated fields in the megagauss  ($\mathord{\sim}100$ tesla) range have been observed to occur near laser hotspots on direct-drive capsule shells  \cite{TimeEvoDirectDrive,SelfGenDirectDrive}, and hohlraum walls  \cite{LiPRL}. These fields have the ability to inhibit thermal transport and raise plasma temperatures \cite{stamper_1991,Farmer2017}. Furthermore, the potential of an externally-imposed field to improve performance and potentially cross the ignition barrier has been demonstrated in magnetised liner inertial fusion (Mag\capsformat{lif}) \cite{MagLIFPoP,MagLIFPRL,MagLIFPreheat,MagLIFExperiment}, 
 plasma-liner-driven magneto-inertial fusion (\capsformat{mif}) \cite{LinerMIFKnapp,MIFRyzhkov,LinerMIFExperiment},
direct-drive \capsformat{icf} \cite{MIFDirectDesign,MIFDirectChang,MIFDirectExperiment} and indirect-drive \capsformat{icf}
 \cite{Montgomery,Strozzi2015,Perkins2017}.

Critical to understanding magnetic field dynamics in laser-plasmas is the Nernst effect. 
Classical (Braginskii \cite{Braginskii}) plasma transport theory shows that this advects magnetic field down temperature gradients at the Nernst velocity
$v_\mathrm{N} = (-)\beta_\wedge \nabla_\perp T_\mathrm{e} /eB$, where $\beta_\wedge$ is a component of the thermoelectric tensor as defined by Epperlein and Haines \cite{EppHaines}, $\vv{\nabla}_\perp T_\mathrm{e}$ is the electron temperature gradient perpendicular to the magnetic field, $e$ is the magnitude of the electron charge and $B$ is the magnitude of the magnetic field. Note that in this paper, the electron temperature is always taken to be in energy units (i.e.\ Boltzmann's constant is taken to be 1). Typically, $v_\mathrm{N}$ lies between the ion sound speed and the electron thermal velocity \cite{willingale}, leading to the build-up of magnetic field at the foot of the temperature gradient, a process known as convective amplification \cite{Nishiguchi}. The consequent cavitation of the magnetic field in hot regions of the plasma degrades its desirable insulating properties \cite{reemerge};
and recent indirect-drive simulations have  demonstrated that neglecting the effect of Nernst advection on
 self-generated fields can lead to a 1.5 \si{\kilo\electronvolt} overestimation of the plasma temperature \cite{Farmer2017}.

However, Davies \etal\ have found that corrections to the classical Nernst velocity (through a flux-limiter) are necessary for matching simulated yield and ion temperature to a direct-drive experiment with an externally imposed field \cite{Davies}. 
This is likely due to nonlocal effects arising near the shock front where the mean free path (mfp) of suprathermal conduction electrons, travelling around four times the thermal velocity ($v_{\mathrm{T}}=\sqrt{T_{\mathrm{e}}/m_{\mathrm{e}}}$ where $m_{\mathrm{e}}$ is the electron mass) can exceed the temperature gradient scalelength $L_{\mathrm{T}} \approx T_{\mathrm{e}} / \lvert \nabla T_{\mathrm{e}} \rvert$. 
The ability of suprathermals to escape steep gradients
leads to non-Maxwellian features in the high-energy tail of  the electron distribution function (\capsformat{edf})
which provides a dominant contribution to both thermal conduction and the Nernst effect
\cite{HainesCanJPhys,HainesV2}.
This explanation is supported in a  recent work by Hill and Kingham \cite{DomNernst},
where a significant reduction of the peak Nernst velocity compared to the Braginskii prediction is observed in a 2D Vlasov-Fokker-Planck simulation of a non-uniformly irradiated \capsformat{ch}-foil.
Additionally, the authors observed an \textit{enhancement} of the Nernst velocity inside the foil where the temperature gradient is relatively flat.
Similar to the phenomena of nonlocal preheat which is important in directly-driven \capsformat{icf} capsules, 
such an effect could not be captured by flux-limiters.

As an alternative to flux-limitation, a number of more advanced models  have been suggested to account for  nonlocal \textit{thermal} transport.
These models are often based on simplifications of the \capsformat{vfp} equation 
\begin{equation}
\frac{\partial f_\mathrm{e}}{\partial t} + \vv{v}\cdot\vv{\nabla} f_\mathrm{e}- \frac{e}{m_\mathrm{e}}\left(\vv{E}+\frac{\vv{v}\times\vv{B}}{c}\right)\cdot \frac{\partial f_\mathrm{e}}{\partial \vv{v}} = C(f_\mathrm{e}),
\end{equation}
for the evolution of the distribution function $f_\mathrm{e}$, where $\vv{v}$ is the electron velocity and $C$ is the operator representing collisions of electrons with themselves and other species.
Examples of nonlocal models include the commonly used \capsformat{snb} model \cite{SNB}
(which we have shown agrees well with kinetic simulations \cite{BrodrickSNB,SherlockSNB}),
the M1 model \cite{Dario}
and many others \cite{NFLF2014,NFLFImplementation,Omotani,LMV,JiClosure,JiHeld,JiHeldZ,JiHeld2017,Izacard,izacard2017,CMG,VDK,WDMSNB}.
While the majority of these models are limited to purely unmagnetised regimes, 
magnetised extensions have been put forward for both the \capsformat{snb} \cite{SNBmag}
and the M1 model \cite{Dariomag}.
However, the accuracy of these has not yet been verified against fully kinetic simulations.
Furthermore, the magnetised \capsformat{snb} multigroup diffusion model does not prescribe any method for calculating nonlocal corrections to the Nernst velocity.
One possible method (described in \cref{TheorBack}) for obtaining an approximation of nonlocal Nernst with nonlocal models designed only for thermal transport has been explored by Lancia \etal\ \cite{LanciaNonlocalNernst} 
and is used as an inspiration for the direction of this paper.
Application of this technique to a nonlocal model very similar to the \capsformat{snb} \cite{riqui2016,FuchsPrivComm} was able to reduce discrepancies between modeling and experiments of a planar laser-solid interaction \cite{LanciaNonlocalNernst,LanciaObservation}.

The role of nonlocality in magnetised transport is not a new idea and has been previously explored in a number of papers published over thirty years ago. Brackbill and Goldman \cite{Brackbill} were the first to demonstrate that the flux-limiting of \emph{all} transport coefficents more accurately captured features predicted by a collisionless \capsformat{pic} code (\capsformat{venus} \cite{VENUS}). Subsequently, Kho and Haines \cite{KhoHainesPRL,*KhoHainesPoP}, used fully-kinetic \capsformat{vfp} simulations to demonstrate that Nernst advection and thermal conduction tend to be flux-limited to a similar degree while the Righi-Leduc heat flow, which provides a bending around field lines, should be limited more strongly. However, they only considered a plasma of moderate 
and uniform 
ionisation ($Z=10$) and neglected anisotropic electron-electron collisions, which can have a significant effect on the Nernst velocity for low-$Z$ plasmas, such as present in the hohlraum gas-fill (see \cref{TheorBack}). Furthermore, they did not compare the time-integrated effect of using a flux-limited hydrodynamic model against \capsformat{vfp} simulations on plasma profiles. Haines \cite{HainesV2} supported this work with a proof that if the electron-ion collision frequency $\nu_\mathrm{ei}$ is assumed to vary articially as $1/v^2$ (while in reality it varies as $\log\varLambda_{\mathrm{ei}}(v)/v^3$) and electron-electron collisions are ignored then the ratio between the Nernst velocity and the perpendicular heat flow is unaffected by nonlocal modifications to the distribution function.  Finally, Luciani \etal\ \cite{LMB}, developed a convolution model for the Nernst velocity and Righi-Leduc heat flow based on simplifications to the quasistatic \capsformat{vfp} equation. Again, this was not tested against a full \capsformat{vfp} code.

In this work, we aim to bring together and expand upon the existing research on nonlocal Nernst effects by comparing \capsformat{vfp} and flux-limited transport approaches at high and low ionisations.
A theoretical overview will first be presented in \cref{TheorBack} to discuss the dominant terms governing the evolution of temperature and magnetic field profiles in the following simulations as well as the implications of common approximations for the Nernst velocity.
We consider a wide range of one-dimensional test problems including relaxation of a temperature ramp with an initially uniform imposed magnetic field in \cref{sec:ramp},
laser heating of nitrogen in \cref{sec:froula} and a lineout from an indirect-drive \capsformat{hydra} simulation in \cref{sec:microdot}.
The main observation is that while both thermal conduction and Nernst advection
can be strongly affected by nonlocality, their ratio is not.
This allows for a simple method of extending a nonlocal thermal transport model (in this case the \capsformat{snb} model \cite{SNB}, which we have explored previously in unmagnetised plasmas \cite{BrodrickSNB,SherlockSNB}) to approximate the Nernst velocity.


\section{Theoretical Background\label{TheorBack}}

For simplicity in this paper, we restrict ourselves to spatial variation in one direction ($x$) only.  In particular, this avoids the possibility of self-generated fields due to the Biermann battery or other anisotropic effects. For a magnetic field pointing solely in the $z$-direction, the evolution of the magnetic field and temperature profile are determined by 
\begin{equation}
\frac{\partial B_z}{\partial t} = -\frac{\partial E_y}{\partial x}, \frac{\partial T_\mathrm{e}}{\partial t} = -\frac{2}{3n_{\mathrm{e}}}\Big(\frac{\partial Q_x}{\partial x}+\frac{E_y}{\mu_0}\frac{\partial B_z}{\partial x}\Big),
\end{equation}
where $E_y$ is the electric field in the $y$-direction and $Q_x$ is the heat flow parallel to the temperature gradient. 
The final term is the Joule heating term, $\vv{E}\cdot \vv{j} $, where the electric current $\vv{j}$ has been substituted with $\vv{\nabla}\times \vv{B} /\mu_0$ using Ampere's law with the displacement current neglected.
In the \capsformat{vfp} simulations presented in this paper the magnetic field gradients are not very steep,
 meaning that this term is not as important as the divergence of the heat flow in determining the plasma temperature.

Both temperature and magnetic field gradients contribute to their own and each other's evolution. This leads to a number of effects, only four of which contribute in our 1D geometry:
\begin{align}
Q_x &= \myannotate[1.6cm]{Thermal Conduction}{-\kappa_\perp \frac{\partial T_\mathrm{e}}{\partial x}} 
                \myannotate{Ettingshausen Effect}{+\frac{\beta_\wedge T_\mathrm{e}}{e \mu_0} \frac{\partial B_z}{\partial x}}, \\
E_y &= \myannotate[1.6cm]{Nernst Advection}{-\frac{\beta_\wedge}{e} \frac{\partial T_\mathrm{e}}{\partial x}} 
               \myannotate{Resistive Diffusion}{-\frac{\alpha_\perp}{e^2 n_\mathrm{e}^2\mu_0}\frac{\partial B_z}{\partial x}},
\end{align}
where $\mu_0$ is the permeability of free space, $\kappa_\perp$ is the perpendicular thermal conductivity, $\alpha_\perp$ is the perpendicular resistivity and $n_\mathrm{e}$ is the electron density.  Again, the weak magnetic field gradients present mean that the 
Ettingshausen effect and resistive diffusion are small corrections
to the Nernst and thermal conduction terms,
  and are therefore not discussed in detail here. 
Additionally, the heat flow perpendicular to both the magnetic field and temperature profile, is given by
\begin{equation}
Q_y =  \myannotate{Righi-Leduc Heat Flow}{-\kappa_\wedge \frac{\partial T_\mathrm{e}}{\partial x}} + 
\myannotate{Peltier \\ Effect}{\frac{\beta_\perp T_\mathrm{e}}{e \mu_0} \frac{\partial B_z}{\partial x}},
\end{equation}
where $\kappa_\wedge$ and $\beta_\perp$ are elements of the thermal conductivity and thermoelectric tensors respectively. As above only the first term is usually dominant in the cases studied here. 

In the local limit, rational polynomial fits for the transport coefficients as a function of magnetisation or Hall parameter $\chi$ have been calculated by both Braginskii \cite{Braginskii} and more accurately by Epperlein and Haines \cite{EppHaines} for varying degrees of ionisation by assuming that the isotropic part of the \capsformat{edf} is Maxwellian. The magnetisation $\chi=\omega_\mathrm{c}\tau_{\mathrm{B}}$ is calculated as the product of the electron cyclotron frequency $\omega_\mathrm{c} = e B /m_\mathrm{e}$ and the collision time
\begin{equation}
\tau_{\mathrm{B}} = 3\sqrt{\frac{\uppi}{2}} \bigg(\frac{4\uppi \epsilon_0}{e^2}\bigg)^2\frac{\sqrt{m_\mathrm{e}}\;T_\mathrm{e}^{3/2}}{4\uppi Z n_\mathrm{e}\log\varLambda_{\mathrm{ei}}},
\end{equation}
where $\epsilon_0$ is the permittivity of free space and  $\log\varLambda_{\mathrm{ei}}$ is the Coulomb logarithm. 
From this the electron-ion mfp can be calculated as $\lambda_{\mathrm{ei}}=v_{\mathrm{T}}\tau_{\mathrm{B}}$.

A commonly used interpolation formula over ionisations for $\kappa_\perp$ 
is to multiply its value in the $Z=\infty$ (Lorentz) limit ($128/3\uppi$)
by a factor $\xi = (Z+0.24)/(Z+4.2)$ \cite{EppShort}.
This approach is also popular as a method to approximate the effect of 
anisotropic electron-electron collisions in some \capsformat{vfp} codes 
(such as \capsformat{spark} \cite{EppShort}, \capsformat{impact} \cite{IMPACT}, \capsformat{aladin} \cite{riqui2016}, \capsformat{impacta} \cite{impacta} and a previous version of K2 \cite{SherlockSNB})
by boosting the electron-ion collision frequency by $1/\xi$.
While giving the correct $Z$-dependence for the perpendicular thermal conductivity at low magnetisation,
this approach can lead to a large overestimate (up to a factor of 2--3) of other transport coefficients
such as $\beta_\wedge$ which determines the Nernst velocity.

\begin{figure}
\begin{tikzpicture}
\begin{axis}[
xlabel = {$Z$},xmin=0,xmax=80,
ylabel = {$\psi = P_\mathrm{e} \beta_\wedge/eB\kappa_\perp$},ymin=0.25,ymax=0.75,
legend pos = south east
]
\addplot[themeGreen,mark=o] table {data/z_ratio_nomag}; \addlegendentry{$\chi \rightarrow 0$}
\addplot[themeOrange,mark=x] table {data/z_ratio_infmag};\addlegendentry{$\chi \rightarrow \infty$}
\addplot[themeGreen,dashed,domain=-1:81,samples=2] {0.7248895};
\addplot[themeOrange,dashed,domain=-1:81,samples=2] {0.461538};
\end{axis}
\end{tikzpicture}
\caption{\label{fig:local_ratio}The local prediction due to Epperlein and Haines \cite{EppHaines} (see  \cref{eq:psiloc}) for the dimensionless quantity  $\psi =P_\mathrm{e}\beta_\wedge /e B \kappa_\perp$ in the limit of zero and infinite magnetisation. 
Dashed lines show the values obtained using the anisotropic collision fix $\xi = (Z+0.24)/(Z+4.2)$ \cite{EppShort}, 
which turn out to be independent of ionisation.}
\end{figure}
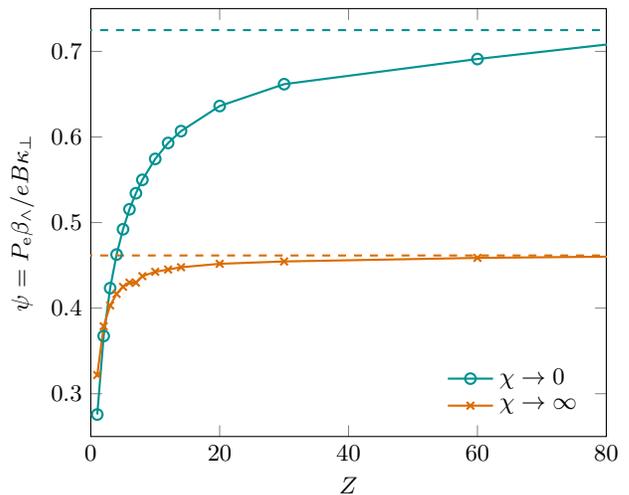

The error induced in the local Nernst velocity by using
the collision fix in \capsformat{vfp} codes instead of
the full anisotropic electron-electron collision operator
can be inferred from \cref{fig:local_ratio}.
This depicts the $Z$-variation  of the dimensionless quantity
$\psi = P_\mathrm{e} \beta_\wedge /eB \kappa_\perp$,
where $P_\mathrm{e}=n_\mathrm{e}T_\mathrm{e}$ is the electron pressure,
in the limit of zero and infinite magnetisation.
The parameter $\psi$ quantifies the ratio between $v_\mathrm{N}$
and the heat flow velocity $v_Q \propto Q_\perp / P_\mathrm{e}$ 
and can be calculated using the polynomial coefficients calculated by Epperlein and Haines
(appearing in table III and IV of their paper \cite{EppHaines}) as
\begin{equation} \label{eq:psiloc}
\lim_{\chi\to 0} \psi = \frac{\beta_{0\wedge}}{\gamma_0},
\lim_{\chi\to \infty} \psi = \frac{(\beta_1''=1.5)}{\gamma_1'}.
\end{equation}
If the collision fix is used as an alternative to fully accounting for electron-electron collisions
(dashed lines) 
then the value of $\psi$ in the zero and infinite magnetisation limits
becomes independent of ionisation and equal to their Lorentz limit values 
(0.46 and 0.73 respectively). 
This is due to cancellation of $\xi$ in the two places it appears: in the $\tau_{\mathrm{B}}$-dependence of $\kappa_\perp$ and the $\chi$-dependence of $\beta_\wedge$.
Consequently, the local Nernst velocity can be greatly overestimated by this approach at low ionisations (by a factor greater than two in for a low magnetisation hydrogen plasma).

Two simple, constant approximations for $\psi$ have been previously suggested and used: 
Nishiguchi \etal\ \cite{Nishiguchi} first suggested that $\psi\approx2/3$,
which obtains the low magnetisation limit to within 10\% for $Z>12$
but overestimates the Nernst velocity at high magnetisation by 40\% or more. 
This was first used by Kho and Haines \cite{KhoHainesPRL,*KhoHainesPoP} 
to demonstrate that the link between Nernst advection and perpendicular heat flow
was not greatly affected by nonlocality,
and more recently by Lancia \etal\ \cite{LanciaNonlocalNernst,riqui2016} 
who used reduced nonlocal heat flow model
to provide a nonlocal prediction of the Nernst velocity as $v_{\mathrm{N}} \approx 2 Q_\perp / 3 P_\mathrm{e} $.
Alternatively, Haines \cite{HainesV2} proved analytically that if the collision frequency is
incorrectly assumed to vary as $1/v^2$ then $\psi = 2/5$ 
even if the isotropic part of the distribution function is far from Maxwellian.
This is the approach employed by Davies \etal\ \cite{Davies}
in calculating the flux-limited Nernst velocity
and works very well for values of $Z$ between 2 and 3,
but provides an underestimate of $\mathord{\sim}1$5--80\% at higher ionisations and low magnetisations.
Due to the potential for large errors arising when treating $\psi$ 
as a constant
we therefore recommend using a fully ionisation- and magnetisation-dependent approach based on the Epperlein and Haines coefficients in both local and reduced nonlocal models.

\section{Temperature Ramp Relaxation\label{sec:ramp}}

\subsection{Methodology}

We have investigated the relaxation of a temperature ramp
in the presence of an initially uniform magnetic field and neglect ion hydrodynamics.
Fully-ionised helium ($Z=2$) and zirconium ($Z=40$) plasmas were studied
to cover the range of ionisations that are typical in hohlraum gas-fill and gold bubble ablation, 
both with fixed and uniform electron densities of
\SI[per-mode=reciprocal]{5d20}{\per\centi\meter\cubed}.
The Coulomb logarithm was taken to be constant at 7.09 in both cases.
The initial temperature profile connecting to the two regions of
\SI{1}{\kilo\electronvolt} and \SI{150}{\electronvolt} respectively and is given by
\begin{equation}
T_{\mathrm{e}}/\si{\electronvolt} = 575-425\tanh(x/L),
\end{equation}
where the initial scalelength $L$ was \SI{50}{\micro\meter} for the helium simulations
and \SI{17.3}{\micro\meter} for the zirconium in order to impose a similar degree of nonlocality.
The simulation domains extended
$\pm7L$, and reflective boundary conditions were used,
restricting heat flow and electric field values to be zero at the boundaries.
A range of initial magnetic fields were considered.
For convenience, we provide a formula to calculate the magnetisation in the hottest and coldest regions of the plasma:
\begin{align}
\mathrlap{\chi^{(1 \mathrm{keV})}}\phantom{\chi^{(150 \mathrm{eV})}} &= 0.54  \times (B_z/\mathrm{tesla})/Z,\\
\chi^{(150 \mathrm{eV})} &= 0.031 \times (B_z/\mathrm{tesla})/Z.
\end{align}

The helium simulations were performed using two \capsformat{vfp} codes---K2  \cite{SherlockSNB} 
and \capsformat{oshun}  \cite{OSHUN2011,OSHUN2013,Joglekar2018}---both 
based on the \capsformat{kalos} formalism \cite{KALOS}. 
This formalism expands the distribution function in spherical harmonics
 and uses a mixture of implicit and explicit time differencing through operator splitting.
 Both codes use the full anisotropic electron-electron collision operator,
 as is necessary to achieve acceptable values of $\beta_\wedge$ at low ionisations,
 even in the local limit.
 Typically, the K2 simulations used spherical harmonics up to order 1,
 and the \capsformat{oshun} simulations up to order 2.
 The codes showed reasonable agreement with each other and
 slight discrepancies were attributable to the number of harmonics used
 and exact implementation of boundary conditions.
 K2 solves for the magnetic and electric fields explicitly using Faraday's law and the Ampere-Maxwell law with an artificial multiplier of \num{100} on the permittivity. 
 This effectively reduces the plasma frequency allowing for larger timesteps of \SI{0.5}{\femto\second}.
 The simulation domain extended from \SI{-350}{\micro\meter} to \SI{350}{\micro\meter} over 100 cells (\SI{7}{\micro\meter} in width) and the uniform velocity grid consisted of 240 cells peaking at \SI{9.4e6}{\meter\per\second} (\SI{25}{\kilo\electronvolt}).

For the zirconium simulations we instead used the fully-implicit code \capsformat{impact} \cite{IMPACT},
which does not include the full collision operator for the angular scattering of electrons with themselves
 in the equation for the first anisotropic part of the \capsformat{edf} $\vv{f}_{\mkern-3mu\relax 1}$.
As a substitute, the electron-ion collision frequency is increased by dividing it by the aforementioned (see \cref{TheorBack}) collision fix $\xi$.
At such high ionisations, the percentage error on the Nernst coefficient $\beta_\wedge$ due to using this approximation is below 10\%.
The advantage of using the collision fix here was the absence of transport coefficients for $Z=40$ in the literature \cite{EppHaines} to compare with the classical transport simulations
(although these could be derived).
In addition to the \capsformat{edf} both the electric and magnetic fields were treated implicity; 
this involved neglecting the displacement current in the Ampere-Maxwell law (see \cite{IMPACT} for more details).
The electron inertia term ($\partial \vv{f}_{\mkern-3mu\relax 1}/ \partial t$) was retained.
The simulation parameters used were a spatial domain extending from $-9L$ to $7L$ over 800 cells (each with a width of \SI{3.46}{\micro\meter}), a uniform velocity grid extending up to \SI{1.8e7}{\meter\per\second} (\SI{94}{\kilo\electronvolt}) and a timestep of \SI{3.35}{\femto\second}.

The distribution functions for the \capsformat{vfp} simulations were initialised as isotropic Maxwellians,
with the anisotropic part (and thereby the heat flow) and electric field initially growing from zero.
Initial transient effects damped within \SI{12}{\pico\second} in the helium and \SI{2}{\pico\second} in the zirconium simulations (about 4 corrected collision times, $\xi \tau_{\mathrm{B}}$, of suprathermal 3--4 keV electrons);
this was determined by both $Q_x$ and $E_y$ reaching a maximum.
We observed that the electric field takes longer to reach its peak than the heat flow,
most likely 
due to the Nernst coefficient $\beta_\wedge$ depending on higher moments of the \capsformat{edf} 
than $\kappa_\perp$  (see the appendix),
making it more sensitive to the dynamics of less collisional high-energy electrons.
The magnetic field and temperature profiles at this point of the simulations (\SI{12}{\pico\second} for helium \SI{2}{\pico\second} for zirconium) were then used to
initialise classical transport simulations with
various combinations of Nernst and thermal flux-limiters.

Our Classical Transport Code (\capsformat{ctc})  \cite{CTC} provides a fully-implicit solution
for the coupled evolution of magnetic field and temperature profiles
using the Epperlein and Haines polynomial fits for the transport coefficients \cite{EppHaines}.
For the zirconium simulation we used the Lorentz limit ($Z=\infty$) transport coefficients 
but multiplied the average collision time $\tau_{\mathrm{B}}$ by the collision fix $\xi$.
The code also has the potential to deal with hydrodynamics and super-Gaussian transport coefficients arising from inverse bremsstrahlung absorption of laser energy\cite{InvBrem,Dum1,*Dum2},
neither of which are used here. Independent Nernst and thermal flux-limiters ($f_\mathrm{N}$ and $f_Q$ respectively)
are available and calculated by multiplying the appropriate transport coefficients 
($\beta_\wedge$ and $\kappa_\perp$)
by a 
spatially-dependent flux-limiting factor $\theta$
(e.g.\ $\kappa_\perp^{\mathrm{(FL)}} = \theta_Q \kappa_\perp^{\mathrm{(Local)}},
\beta_\wedge^{\mathrm{(FL)}} = \theta_{\mathrm{N}} \beta_\wedge^{\mathrm{(Local)}},$)
which always depends on the ratio of perpendicular local thermal conduction to the free-streaming limit $Q_\mathrm{fs} =  v_\mathrm{T}  P_\mathrm{e}$ (where $v_\mathrm{T}=\sqrt{T_\mathrm{e}/m_\mathrm{e}}$):
\begin{equation}
\label{eq:fluxlim}
\theta_{\alpha} = \bigg(1+\frac{\kappa_\perp}{f_{\alpha} Q_\mathrm{fs}} \bigg\lvert\frac{\partial T_\mathrm{e}}{\partial x}\bigg\rvert\bigg)^{\smash{-1}},
\end{equation}
where $\alpha = {\mathrm{N}}$ or $Q$ \cite{CTCfluxlim}.
We believe this definition of $f$ to be consistent 
with that used by popular \capsformat{icf} hydro codes  (e.g.\ \capsformat{hydra}, \capsformat{lasnex} \cite{HYDRA}, \capsformat{lilac} and \capsformat{draco} \cite{Davies}),
but these codes typically limit the heat flow
to the minimum of $Q_\mathrm{fs}$ and $Q_\perp$
rather than half their harmonic average as presented here.
We also present results for the flux-limited Righi-Leduc heat flow
which are obtained in the same way,
i.e.\ by multiplying $\kappa_\wedge$ by $\theta_\mathrm{RL}$ using an independent Righi-Leduc flux limiter $f_\mathrm{RL}$.

\subsection{Results}

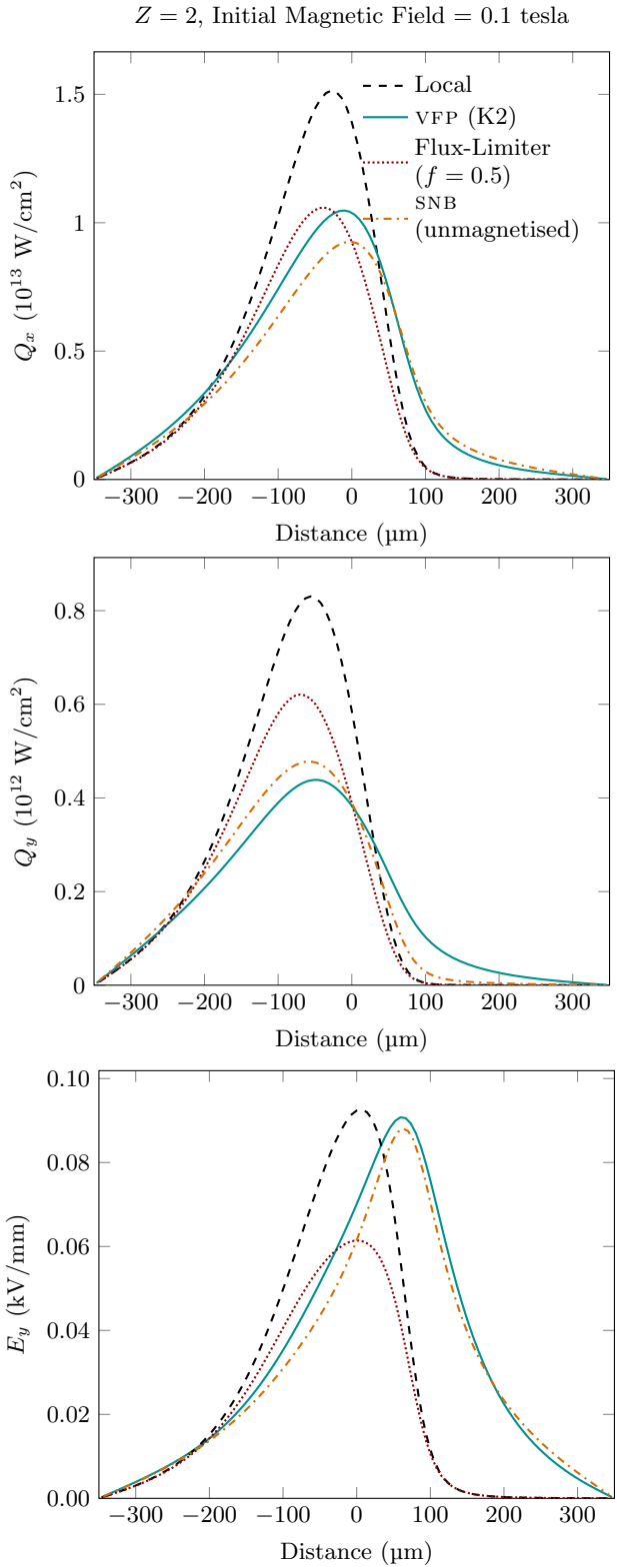
\begin{figure}[p]
\begin{tikzpicture}
\begin{axis}[
title = {$Z=2$, Initial Magnetic Field = 0.1 tesla},
xlabel = {Distance (\si{\micro\metre})},xmin=-350.,xmax=350.,
ylabel = {$Q_x$ ($10^{13}$ \si{\watt\per\centi\meter\squared})},ymin=0.,
scaled y ticks={real:1e13},
ytick scale label code/.code={},
legend pos = north east,
]
\addplot[local] table {data/K2New_0.1Tesla_local_xmic_QxWcm2_15.0ps}; \addlegendentry{Local}
\addplot[vfp] table {data/K2New_0.1Tesla_k2_xmic_QxWcm2_15.0ps}; \addlegendentry{\capsformat{vfp} (K2)}
\addplot[fluxlim] table {data/K2New_0.1Tesla_rf2.8284rN2.8284_xmic_QxWcm2_15.0ps}; \addlegendentry{Flux-Limiter \\  ($f=0.5$)}
\addplot[snb] table {data/K2New_0.1Tesla_snb_xmic_QxWcm2_15.0ps}; \addlegendentry{\capsformat{snb} \\ (unmagnetised)}
\end{axis}
\end{tikzpicture}

\begin{tikzpicture}
\begin{axis}[
xlabel = {Distance (\si{\micro\metre})},xmin=-350.,xmax=350.,
ylabel = {$Q_y$ ($10^{12}$ \si{\watt\per\centi\meter\squared})},ymin=0.,
scaled y ticks={real:1e12},
ytick scale label code/.code={},
legend pos = north west,
]
\addplot[local] table {data/K2New_0.1Tesla_local_xmic_QyWcm2_15.0ps};
\addplot[vfp] table {data/K2New_0.1Tesla_k2_xmic_QyWcm2_15.0ps};
\addplot[fluxlim] table {data/K2New_0.1Tesla_fluxlimperp_xmic_QyWcm2_15.0ps};
\addplot[snb] table {data/K2New_0.1Tesla_snb_xmic_QyWcm2_15.0ps};
\end{axis}
\end{tikzpicture}

\begin{tikzpicture}
\begin{axis}[
xlabel = {Distance (\si{\micro\metre})},xmin=-350.,xmax=350.,
ylabel = {$E_y$ (\si{\kilo\volt\per\milli\metre})},ymin=0.,
    y tick label style={
        /pgf/number format/.cd,
            fixed,
            fixed zerofill,
            precision=2,
        /tikz/.cd
    },
]
\addplot[vfp] table {data/K2New_0.1Tesla_k2_xmic_EyMV_15.0ps};
\addplot[local] table {data/K2New_0.1Tesla_local_xmic_EyMV_15.0ps};
\addplot[fluxlim] table {data/K2New_0.1Tesla_rf2.8284rN2.8284_xmic_EyMV_15.0ps};
\addplot[snb] table {data/K2New_0.1Tesla_snb_xmic_EyMV_15.0ps};
\end{axis}
\end{tikzpicture}
\caption{\label{fig:k2_0.1tesla} Perpendicular and Righi-Leduc heat flows $Q_x$ (top), $Q_y$ (middle) and the Nernst-dominated out-of-plane electric field $E_y$ (bottom) after 15 ps K2 \capsformat{vfp} helium simulation with an initial magnetic field of 0.1 tesla. Local, flux-limited and \capsformat{snb} profiles were postprocessed using the K2 temperature and magnetisation profiles. \capsformat{snb} $E_y$ and $Q_y$ are calculated by multiplying the (unmagnetised) \capsformat{snb}  $Q_x$ profile by the corresponding ratio in the local limit ($E_y/Q_x, Q_y/Q_x$).}
\end{figure}

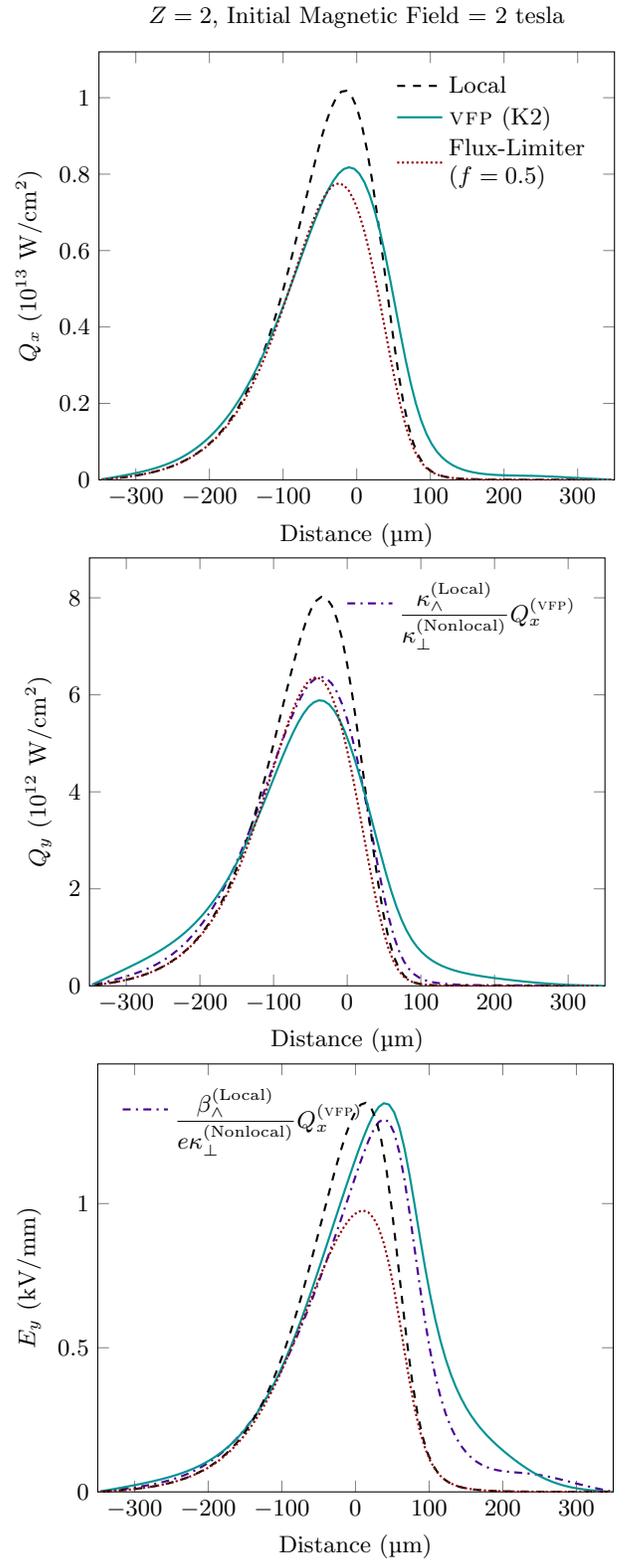
\begin{figure}[p]
\begin{tikzpicture}
\begin{axis}[
title = {$Z=2$, Initial Magnetic Field = 2 tesla},
xlabel = {Distance (\si{\micro\metre})},xmin=-350.,xmax=350.,
ylabel = {$Q_x$ ($10^{13}$ \si{\watt\per\centi\metre\squared})},ymin=0.,
scaled y ticks={real:1e13},
ytick scale label code/.code={},
legend pos = north east
]
\addplot[local] table {data/K2New_2.0Tesla_local_xmic_QxWcm2_12.0ps}; \addlegendentry{Local}
\addplot[vfp] table {data/K2New_2.0Tesla_k2_xmic_QxWcm2_12.0ps}; \addlegendentry{\capsformat{vfp} (K2)}
\addplot[fluxlim] table {data/K2New_2.0Tesla_rf2.8284_rN2.8284_xmic_QxWcm2_12.0ps}; \addlegendentry{Flux-Limiter \\  ($f=0.5$)}
\end{axis}
\end{tikzpicture}

\begin{tikzpicture}
\begin{axis}[
xlabel = {Distance (\si{\micro\metre})},xmin=-350.,xmax=350.,
ylabel = {$Q_y$ ($10^{12}$ \si{\watt\per\centi\metre\squared})},ymin=0.,
scaled y ticks={real:1e12},
ytick scale label code/.code={},
legend pos = north east,
]
\addplot[haines] table {data/K2New_2.0Tesla_Hainesk2_xmic_QyWcm2_12.0ps}; 
 \addlegendentry{$\displaystyle\frac{\kappa_\wedge^{\mathrm{(Local)}}}{\kappa_\perp^{\mathrm{(Nonlocal)}}}Q_x^{(\capsformat{vfp})}$}
\addplot[local] table {data/K2New_2.0Tesla_local_xmic_QyWcm2_12.0ps}; 
\addplot[vfp] table {data/K2New_2.0Tesla_k2_xmic_QyWcm2_12.0ps}; 
\addplot[fluxlim] table {data/K2New_2.0Tesla_fluxlimperp_xmic_QyWcm2_12.0ps}; 
\end{axis}
\end{tikzpicture}

\begin{tikzpicture}
\begin{axis}[
xlabel = {Distance (\si{\micro\metre})},xmin=-350.,xmax=350.,
ylabel = {$E_y$ (\si{\kilo\volt\per\milli\metre})},ymin=0.,
legend pos = north west,
]
\addplot[haines] table {data/K2New_2.0Tesla_Hainesk2_xmic_EyMV_12.0ps};  \addlegendentry{$\displaystyle\frac{\beta_\wedge^{\mathrm{(Local)}}}{e\kappa_\perp^{\mathrm{(Nonlocal)}}}Q_x^{(\capsformat{vfp})}$}
\addplot[vfp] table {data/K2New_2.0Tesla_k2_xmic_EyMV_12.0ps};
\addplot[local] table {data/K2New_2.0Tesla_local_xmic_EyMV_12.0ps};
\addplot[fluxlim] table {data/K2New_2.0Tesla_rf2.8284_rN2.8284_xmic_EyMV_12.0ps};
\end{axis}
\end{tikzpicture}
\caption{\label{fig:k2_2tesla}Perpendicular heat flow $Q_x$ (top), $Q_y$ (middle) and the Nernst-dominated out-of-plane electric field $E_y$ (bottom) after 12 ps K2 \capsformat{vfp} helium simulation with an initial magnetic field of 2 tesla. Local and flux-limited profiles were postprocessed using the K2 temperature and magnetisation profiles.
}
\end{figure}

\begin{figure}[p]
\begin{tikzpicture}
\begin{axis}[
title = {$Z=40$, Initial Magnetic Field = 1 tesla},
xlabel = {Distance (\si{\micro\metre})},xmin=-125.,xmax=125.,
ylabel = {$Q_x$ ($10^{12}$ \si{\watt\per\centi\meter\squared})},ymin=0.,
scaled y ticks={real:1e12},
ytick scale label code/.code={},
legend pos = north east,
]
\addplot[local] table {data/z40_35.5mic_1.tesla_noinit_local_xmic_QxWcm2_04.0ps}; \addlegendentry{Local}
\addplot[vfp] table {data/z40_35.5mic_1.tesla_noinit_impact_xmic_QxWcm2_04.0ps}; \addlegendentry{\capsformat{vfp} (\capsformat{impact})}
\addplot[fluxlim] table {data/z40_35.5mic_1.tesla_noinit_f.15_xmic_QxWcm2_04.0ps}; \addlegendentry{Flux-Limiter \\  ($f=0.15$)}
\addplot[snb] table {data/z40_35.5mic_1.tesla_noinit_snb_xmic_QxWcm2_04.0ps}; \addlegendentry{\capsformat{snb} \\ (unmagnetised)}
\end{axis}
\end{tikzpicture}

\begin{tikzpicture}
\begin{axis}[
xlabel = {Distance (\si{\micro\metre})},xmin=-125.,xmax=125.,
ylabel = {$Q_y$ ($10^{12}$ \si{\watt\per\centi\meter\squared})},ymin=0.,
scaled y ticks={real:1e12},
ytick scale label code/.code={},
legend pos = north west,
]
\addplot[local] table {data/z40_35.5mic_1.tesla_noinit_local_xmic_QyWcm2_04.0ps}; 
\addplot[vfp] table {data/z40_35.5mic_1.tesla_noinit_impact_xmic_QyWcm2_04.0ps}; 
\addplot[fluxlim] table {data/z40_35.5mic_1.tesla_noinit_f.15_xmic_QyWcm2_04.0ps}; 
\addplot[snb] table {data/z40_35.5mic_1.tesla_noinit_snb_xmic_QyWcm2_04.0ps}; 
\end{axis}
\end{tikzpicture}

\begin{tikzpicture}
\begin{axis}[
xlabel = {Distance (\si{\micro\metre})},xmin=-125.,xmax=125.,
ylabel = {$E_y$ (\si{\kilo\volt\per\milli\metre})},ymin=0.,
    y tick label style={
        /pgf/number format/.cd,
            fixed,
            fixed zerofill,
            precision=2,
        /tikz/.cd
    },
]
\addplot[local] table {data/z40_35.5mic_1.tesla_noinit_local_xmic_EykVmm_04.0ps}; 
\addplot[vfp] table {data/z40_35.5mic_1.tesla_noinit_impact_xmic_EykVmm_04.0ps}; 
\addplot[fluxlim] table {data/z40_35.5mic_1.tesla_noinit_f.15_xmic_EykVmm_04.0ps}; 
\addplot[snb] table {data/z40_35.5mic_1.tesla_noinit_snb_xmic_EykVmm_04.0ps}; 
\end{axis}
\end{tikzpicture}
\caption{\label{fig:z40_1tesla} Perpendicular and Righi-Leduc heat flows $Q_x$ (top), $Q_y$ (middle) and the Nernst-dominated out-of-plane electric field $E_y$ (bottom) after 4 ps \capsformat{impact}\ \capsformat{vfp} zirconium simulation with an initial magnetic field of 1 tesla. Local, flux-limited and \capsformat{snb} profiles were postprocessed using the \capsformat{impact} temperature and magnetisation profiles. \capsformat{snb} $E_y$ and $Q_y$ are calculated by multiplying the (unmagnetised) \capsformat{snb} $Q_x$ profile by the corresponding ratio in the local limit ($E_y/Q_x, Q_y/Q_x$).}
\end{figure}
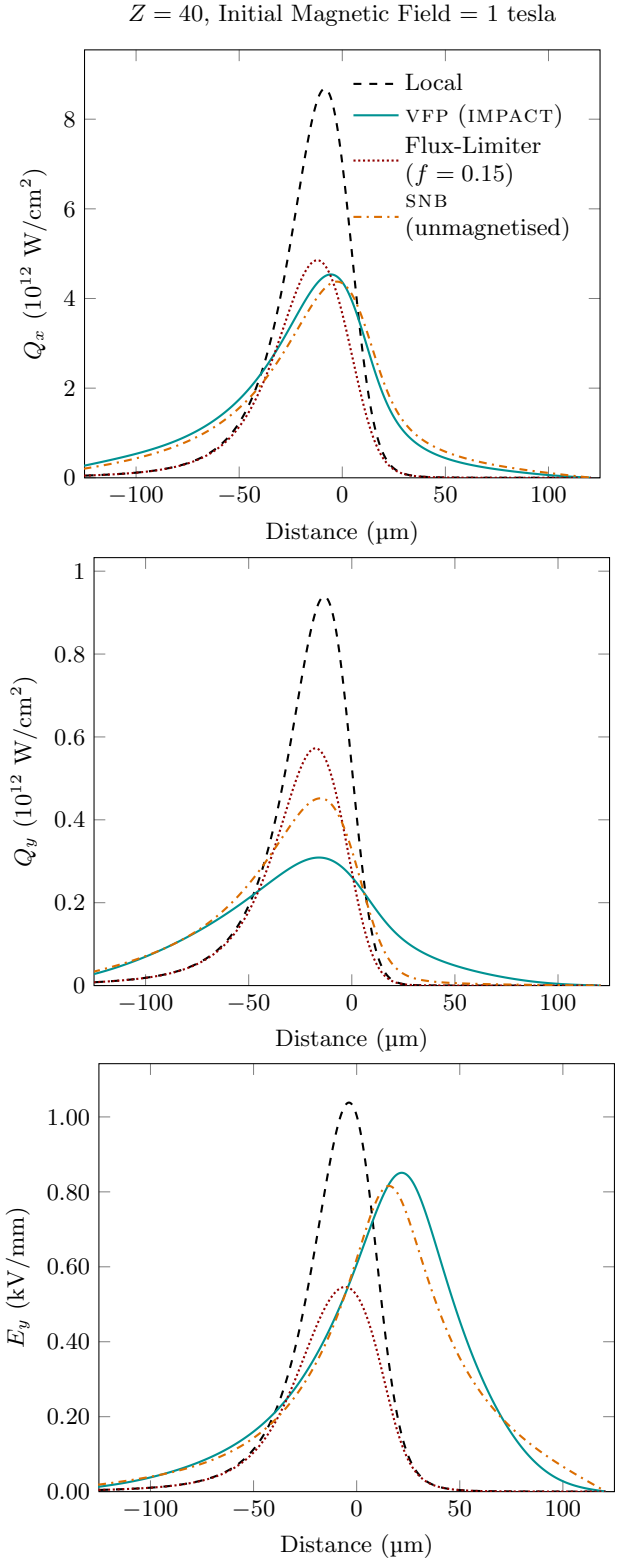

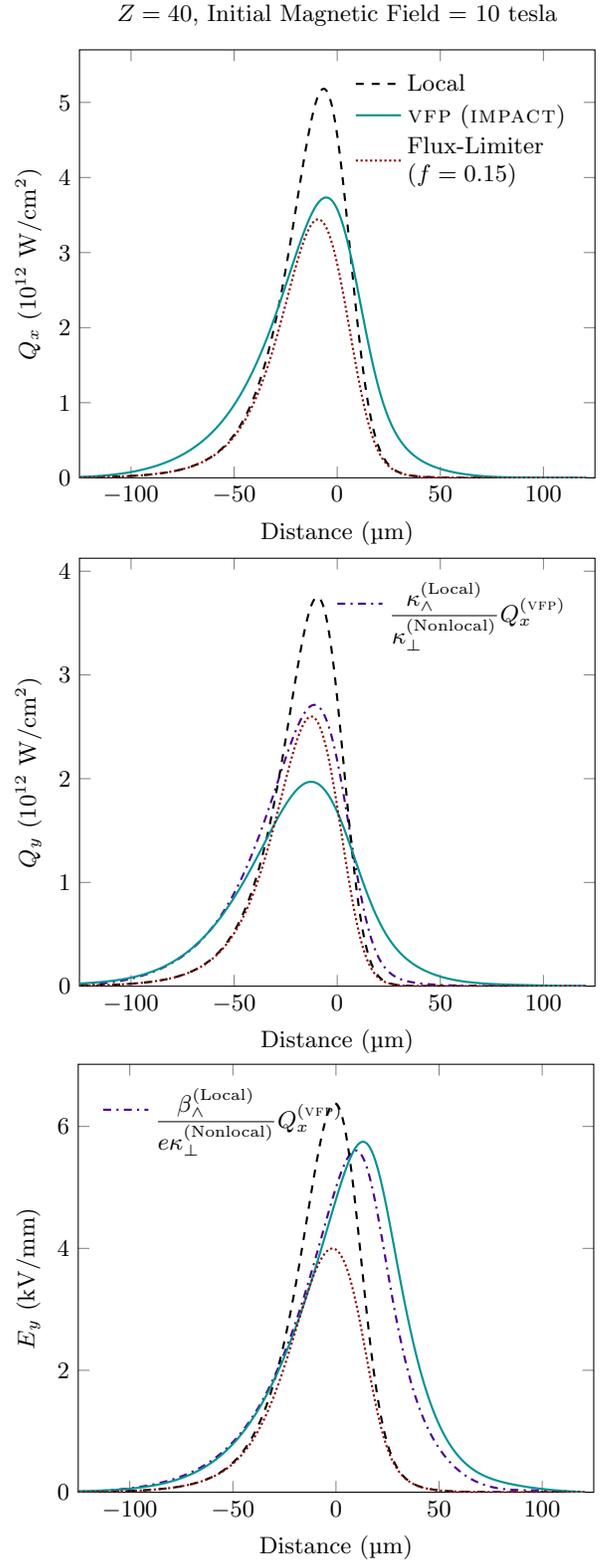
\begin{figure}[p]
\begin{tikzpicture}
\begin{axis}[
title = {$Z=40$, Initial Magnetic Field = 10 tesla},
xlabel = {Distance (\si{\micro\metre})},xmin=-125.,xmax=125.,
ylabel = {$Q_x$ ($10^{12}$ \si{\watt\per\centi\metre\squared})},ymin=0.,
scaled y ticks={real:1e12},
ytick scale label code/.code={},
legend pos = north east
]
\addplot[local] table {data/z40_35.5mic_10.tesla_noinit_local_xmic_QxWcm2_04.0ps}; \addlegendentry{Local}
\addplot[vfp] table {data/z40_35.5mic_10.tesla_noinit_impact_xmic_QxWcm2_04.0ps}; \addlegendentry{\capsformat{vfp} (\capsformat{impact})}
\addplot[fluxlim] table {data/z40_35.5mic_10.tesla_noinit_f.15_xmic_QxWcm2_04.0ps}; \addlegendentry{Flux-Limiter \\  ($f=0.15$)}
\end{axis}
\end{tikzpicture}

\begin{tikzpicture}
\begin{axis}[
xlabel = {Distance (\si{\micro\metre})},xmin=-125.,xmax=125.,
ylabel = {$Q_y$ ($10^{12}$ \si{\watt\per\centi\metre\squared})},ymin=0.,
scaled y ticks={real:1e12},
ytick scale label code/.code={},
legend pos = north east,
]
\addplot[haines] table {data/z40_35.5mic_10.tesla_noinit_haines_xmic_QyWcm2_04.0ps};  \addlegendentry{$\displaystyle\frac{\kappa_\wedge^{\mathrm{(Local)}}}{\kappa_\perp^{\mathrm{(Nonlocal)}}}Q_x^{(\capsformat{vfp})}$}
\addplot[local] table {data/z40_35.5mic_10.tesla_noinit_local_xmic_QyWcm2_04.0ps}; 
\addplot[vfp] table {data/z40_35.5mic_10.tesla_noinit_impact_xmic_QyWcm2_04.0ps}; 
\addplot[fluxlim] table {data/z40_35.5mic_10.tesla_noinit_f.15_xmic_QyWcm2_04.0ps}; 
\end{axis}
\end{tikzpicture}

\begin{tikzpicture}
\begin{axis}[
xlabel = {Distance (\si{\micro\metre})},xmin=-125.,xmax=125.,
ylabel = {$E_y$ (\si{\kilo\volt\per\milli\metre})},ymin=0.,
legend pos = north west,
]
\addplot[haines] table {data/z40_35.5mic_10.tesla_noinit_haines_xmic_EykVmm_04.0ps};  \addlegendentry{$\displaystyle\frac{\beta_\wedge^{\mathrm{(Local)}}}{e\kappa_\perp^{\mathrm{(Nonlocal)}}}Q_x^{(\capsformat{vfp})}$}
\addplot[local] table {data/z40_35.5mic_10.tesla_noinit_local_xmic_EykVmm_04.0ps}; 
\addplot[vfp] table {data/z40_35.5mic_10.tesla_noinit_impact_xmic_EykVmm_04.0ps}; 
\addplot[fluxlim] table {data/z40_35.5mic_10.tesla_noinit_f.15_xmic_EykVmm_04.0ps}; 
\end{axis}
\end{tikzpicture}
\caption{\label{fig:z40_10tesla}Perpendicular and Righi-Leduc heat flows $Q_x$ (top), $Q_y$ (middle) and the Nernst-dominated out-of-plane electric field $E_y$ (bottom) after 4 ps \capsformat{impact} \capsformat{vfp} zirconium simulation with an initial magnetic field of 10 tesla. Local and flux-limited profiles were postprocessed using the \capsformat{impact} temperature and magnetisation profiles. 
}
\end{figure}

Instantaneous snapshots of perpendicular heat flow ($Q_x$), Righi-Leduc heat flow ($Q_y$) and the Nernst-relevant out-of-plane electric field ($E_y$) at the end of the initial transient periods are respectively presented in the top, middle and bottom panels of  \cref{fig:k2_0.1tesla,fig:k2_2tesla,fig:z40_10tesla,fig:z40_1tesla} for selected simulations:
Low and high magnetisation helium runs are presented in \cref{fig:k2_0.1tesla,fig:k2_2tesla} corresponding to initial magnetic fields of 0.1 tesla and 2 tesla respectively.
For the zirconium runs presented in \cref{fig:z40_10tesla,fig:z40_1tesla} we provide profiles resulting from initial magnetic fields of 1 tesla and 10 tesla respectively.
As results from the various simulations are qualitatively similar we shall discuss them all simultaneously.
We shall first compare these profiles to what would be predicted using the Epperlein and Haines transport theory before considering the instantaneous and time-integrated effects of using flux-limiters. 
Finally, in \cref{snb}, we shall outline and examine the possibility of using a more sophisticated reduced model such as \capsformat{snb}.

With respect to the local predictions it is clear that there are both flux-reduction and preheat effects in the $Q_x$ and $Q_y$ heat flow profiles, 
but these nonlocal effects are more pronounced for the Righi-Leduc heat flow due to its dependence on higher moments of the distribution function (see \cref{app:mom}).
On the other hand, the electric field mainly experiences a shift in the peak toward the cooler region of the plasma
with little reduction in its actual value 
and in fact an \textit{increase} of the value of the electric field in the colder region of the plasma where the temperature gradient is relatively flat,
 which we shall here refer to as `pre-Nernst.'
 These observations are qualitatively similar to those previously seen by both Kho and Haines \cite{KhoHainesPRL,*KhoHainesPoP} and Hill and Kingham \cite{DomNernst}.

It may seem surprising that, despite the similar degrees of nonlocality and relative flux-limitation of the helium and zirconium simulations,
the actual values of flux-limiters 
 deemed optimal (by eye) for $Q_x$ turn out to be quite different
(0.5 for helium and 0.15 for zirconium).
However,
this is simply due to differences in the  $Z$-dependence of the perpendicular thermal conductivity $\kappa_\perp \propto  \xi / Z$ and the nonlocality parameter $\sqrt{\xi Z} \lambda_{\mathrm{ei}} \propto \sqrt{\xi / Z}$.
(The appearance of the multiplier $\sqrt{Z}$ in the nonlocality parameter dates back at least as far as seminal work by Luciani, Mora and Virmont \cite{LMV}, and the later incorporation of the collision fix can be traced back to Epperlein and Short \cite{EppShort}. 
We additionally refer the reader to  section IV A of our recent paper \cite{BrodrickSNB}, which expands upon original linearised analysis by Bychenkov \etal\  \cite{Bychenkov,*Bychenkov95}, for further discussion.)
In our simulations, we arranged for the maximum nonlocality parameter to be approximately equal to 0.1 for both the helium and zirconium simulations by using different length scales 
$L \propto \sqrt{\xi/Z}$.
Therefore, in order to obtain equivalent flux-limiting factors $\theta$ (see \cref{eq:fluxlim}),
the flux-limiters $f$ need to make up a further factor $\sqrt{\xi/Z}$ to
 fully compensate the ionisation-scaling of the thermal conductivity,
  explaining the discrepancy in their optimal values: $\sqrt{40/\xi(40)}0.15\approx \sqrt{2/\xi(2)} 0.5$.
If we had not used different values for $L$ the values of the flux-limiters would have indeed been similar to each other,
but the resulting flux-limitation factors $\theta$ would be closer to unity for zirconium than helium.	
These observations suggest that it is worth carefully considering
 whether the value of flux-limiter used in laser-plasma codes should be material dependent perhaps through an inline calculation of the nonlocality parameter at each point in space.

As magnetic fields should in theory relocalise the transport it may seem surprising that the optimal flux-limiter value does not appear
to depend greatly on magnetisation.
Nevertheless, the heat flow does indeed approach its local value as higher magnetisations are reached.
This is possible 
because of the reduction of the \textit{local} heat flow in the presence of strong magnetic fields to well below the free-streaming limit.

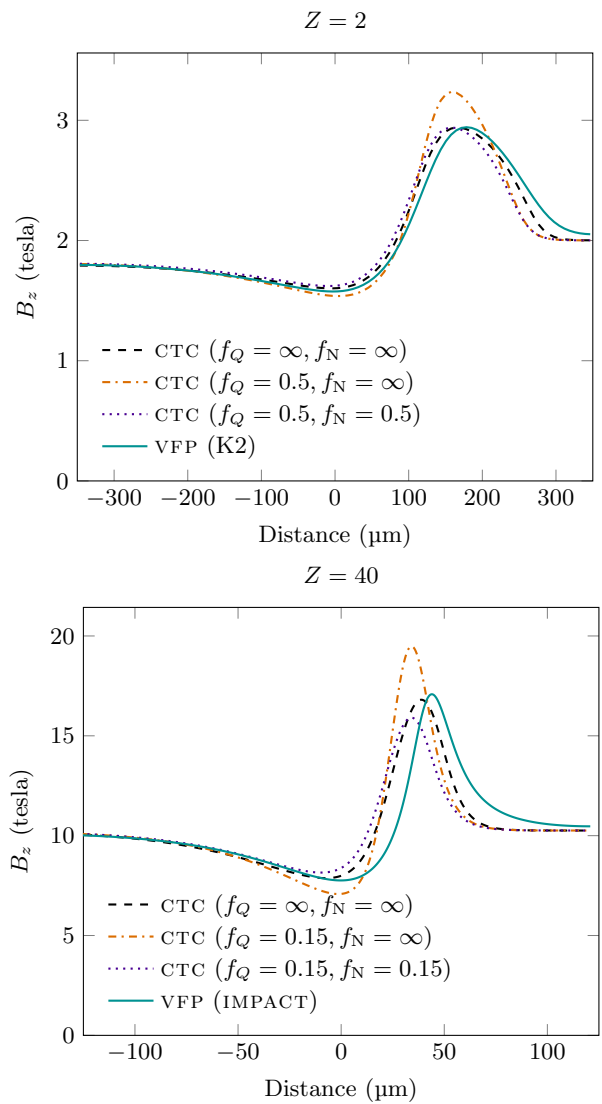
\begin{figure}[t]
\begin{tikzpicture}
\begin{axis}[
title = {$Z=2$},
xlabel = {Distance (\si{\micro\metre})},xmin=-350.,xmax=350.,
ylabel = {$B_z$ (tesla)},ymin=0.,
legend pos = south west,
]
\addplot[local] table {data/K2New_2.0Tesla_nofluxlim_xmic_Btesla_200.0ps}; \addlegendentry{\capsformat{ctc} ($f_Q=\infty, f_\mathrm{N}=\infty$)}
\addplot[snb] table {data/K2New_2.0Tesla_rf2.8284_rN0_xmic_Btesla_200.0ps}; \addlegendentry{\capsformat{ctc} ($f_Q=0.5, f_\mathrm{N}=\infty$)}
\addplot[themePurple,dotted] table {data/K2New_2.0Tesla_rf2.8284_rN2.8284_xmic_Btesla_200.0ps}; \addlegendentry{\capsformat{ctc} ($f_Q=0.5, f_\mathrm{N}=0.5$)}
\addplot[vfp] table {data/K2New_2.0Tesla_k2_xmic_Btesla_200.0ps}; \addlegendentry{\capsformat{vfp} (K2)}
\end{axis}
\end{tikzpicture}
\begin{tikzpicture}
\begin{axis}[
title = {$Z=40$},
xlabel = {Distance (\si{\micro\metre})},xmin=-125.,xmax=125.,
ylabel = {$B_z$ (tesla)},ymin=0.,
legend pos = south west,
]
\addplot[local] table {data/z40_35.5mic_10.tesla_noinit_long_CTCnofluxlim_xmic_BzTesla_50.2ps}; \addlegendentry{\capsformat{ctc} ($f_Q=\infty, f_\mathrm{N}=\infty$)}
\addplot[snb] table {data/z40_35.5mic_10.tesla_noinit_long_CTCrf9.428rN0_xmic_BzTesla_50.2ps}; \addlegendentry{\capsformat{ctc} ($f_Q=0.15, f_\mathrm{N}=\infty$)}
\addplot[themePurple,dotted] table {data/z40_35.5mic_10.tesla_noinit_long_CTCrf9.428rN9.428_xmic_BzTesla_50.2ps}; \addlegendentry{\capsformat{ctc} ($f_Q=0.15, f_\mathrm{N}=0.15$)}
\addplot[vfp] table {data/z40_35.5mic_10.tesla_noinit_long_impact_xmic_BzTesla_50.2ps}; \addlegendentry{\capsformat{vfp} (\capsformat{impact})}
\end{axis}
\end{tikzpicture}
\caption{\label{fig:Bfluxlim}
Comparison of magnetic field profiles predicted by the Classical Transport Code \capsformat{ctc} with different combinations of thermal and Nernst flux-limiters $f_Q, f_\mathrm{N}$ respectively. 
Helium profiles (top) were evolved independently for a further \SI{188}{\pico\second} starting from the K2 $T_\mathrm{e}$ and $B_z$ profiles  at \SI{12}{\pico\second},
while the zirconium profiles (bottom) were simulated independently for a further \SI{48}{\pico\second} from the \capsformat{impact} profiles at \SI{2}{\pico\second}.}
\end{figure}

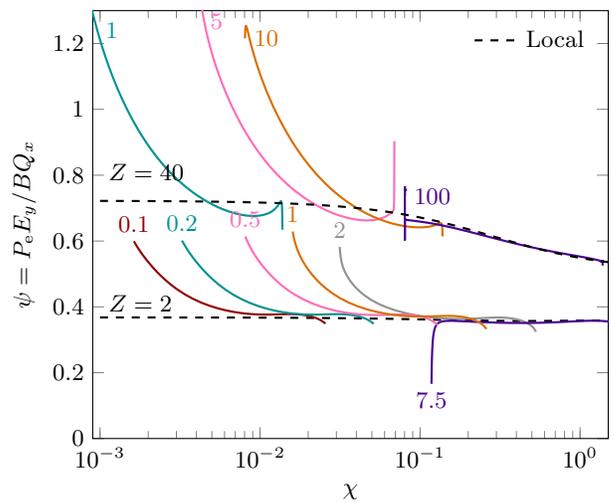
\begin{figure}[t]
\begin{tikzpicture}
\begin{semilogxaxis}[
xlabel = {$\chi$},xmin=0.0009,xmax=1.5,
ylabel = {$\psi = P_\mathrm{e} E_y/ B Q_x $},ymin=0.,ymax=1.3
]
\addplot[local] table {data/local_xi_ratio}; \addlegendentry{Local}

\addplot[themeRed] table {data/K2New_0.1Tesla_k2_chi_ratio_25.0ps} node [pos=1.0,above] {0.1};
\addplot[themeGreen] table {data/K2New_0.2Tesla_k2_chi_ratio_25.0ps} node [pos=1.0,above] {0.2};
\addplot[themeGray] table {data/K2New_2.0Tesla_k2_chi_ratio_25.0ps} node [pos=1.0,above] {2};
\addplot[themePurple] table {data/K2New_7.5Tesla_k2_chi_ratio_25.0ps} node [pos=1.0,below] {7.5};

\addplot[themeGreen] table {data/z40_35.5mic_1.tesla_noinit_IMPACT_chi_ratio_05.0ps} node [pos=0.95,anchor=north west,xshift=1 ex] {1};
\addplot[themePink] table {data/z40_35.5mic_5.tesla_noinit_IMPACT_chi_ratio_05.0ps} node [pos=0.9,anchor=north west] {5};
\addplot[themeOrange] table {data/z40_35.5mic_10.tesla_noinit_IMPACT_chi_ratio_05.0ps} node [anchor=west] {10};
\addplot[themePurple] table {data/z40_35.5mic_100.tesla_noinit_IMPACT_chi_ratio_05.0ps} node [anchor=west] {100};
\node [anchor=west] at (0.001,0.41) {$Z=2$};
\node [anchor=west] at (0.001,0.81) {$Z=40$};

\addplot[local] table {data/local_chi_ratio_z40};

\addplot[themePink] table {data/K2New_0.5Tesla_k2_chi_ratio_25.0ps} node [pos=1.0,above] {0.5};
\addplot[themeOrange] table {data/K2New_1.0Tesla_k2_chi_ratio_25.0ps} node [pos=1.0,above] {1};
\end{semilogxaxis}
\end{tikzpicture}
\caption{\label{fig:ratio_nonlocal}
Solid coloured lines show the variation of  
$\psi = P_\mathrm{e} E_y/ B Q_x $ (proportional to the ratio of nonlocal Nernst and heat flow velocities)
with magnetisation $\chi$
after \SI{25}{\pico\second} \capsformat{vfp} simulation for the helium runs (bottom) and \SI{5}{\pico\second} for the zirconium (top).
Colors differentiate between values of the initial magnetic field, which are labelled in units of tesla.
Dashed lines show the prediction for $\psi$ in the local limit.
The proximity of the centre of the colored lines to the dashed shows that the ratio between the peak magnetic and electric field is not strongly affected by nonlocality (and is in fact more affected by ionisation).
The 50\% overestimate of the local prediction at low magnetisations shows that the pre-Nernst advecting magnetic field beyond the temperature gradient is more pronounced than preheat. 
At higher magnetisations, nonlocality is unimportant at such early times and the observed dip in the value of $\psi$ for the \SI{7.5}{\tesla} run at low temperatures is simply a numerical feature due to the small values of $E_y$ and $Q_x$ in these regions.}
\end{figure}

Looking at the  $E_y$ profiles which determine the evolution of the magnetic field due to Nernst advection,
we see that using the same flux-limiter as for $Q_x$ gets the profile about right in the hot region of the plasma.
However, doing so grossly underestimates the \textit{peak} electric field and 
the flux-limiter approach inherently fails to capture any of the prominent pre-Nernst observed.
The former observation in particular suggests that perhaps it would be desirable to use a larger Nernst flux-limiter in order to match the peak.
Contrastingly, it is clear that a \textit{lower} flux-limiter is necessary on the Righi-Leduc heat flow to capture its higher degree of flux suppression due to nonlocality.

The results shown in \cref{fig:k2_0.1tesla,fig:k2_2tesla,fig:z40_10tesla,fig:z40_1tesla} are sufficiently early in the simulations that the magnetic field profile has not evolved significantly.
The top panel of \cref{fig:Bfluxlim} shows the magnetic field profiles 
predicted by K2 and \capsformat{ctc} with 
varying combinations of flux-limiters at \SI{200}{\pico\second} 
for a helium simulation with an initial magnetic field of \SI{2}{\tesla}
(recall that the classical transport simulation was started \SI{12}{\pico\second} in)
and a similar comparison between \capsformat {impact} and \capsformat{ctc}
for the \SI{10}{\tesla} zirconium at \SI{50}{\pico\second} (with \capsformat{ctc} starting at \SI{2}{\pico\second})
 in the bottom panel.
It is observed that when using only a thermal flux-limiter (orange dash-dotted), as is traditional ,
the relative amplification of the magnetic field is overestimated, by over 30\% for the case of the zirconium simulation.
Additionally, the degree of magnetic cavitation is also slightly overestimated by the traditional approach. 
Incorporating a Nernst limiter noticeably improves agreement with \capsformat{vfp}
but does not account for the smearing and shifting of the peak beyond the foot of the temperature gradient.
Despite these noticeable differences in the final magnetic field profiles these were not sufficient to cause distinguishable modifications on the final temperature profiles. 

Although Nernst advection in the zirconium simulation might be considered slightly over-constrained by a limiter of 0.15,
we still suggest that the most sensible method of limiting Nernst advection is to always use $f_{\mathrm{N}} =f_Q$  as other choices are ad hoc and cannot be justified physically.
This also conveniently prevents the undesirable introduction of an additional tunable parameter.
Also note that when instead dispensing with flux-limiters completely in these simulations 
(i.e.\ $f_Q = f_{\mathrm{N}}=\infty$) 
we achieve the best agreement with the \capsformat{vfp} magnetic field profiles as cold plasma is allowed to heat up quicker, thereby enhancing the spread of magnetic field.
However,
it would indeed be preferable to go beyond flux-limiters to a more predictive approach, 
such as a reduced nonlocal model,
that could account for the prominent smearing and delocalisation effects of pre-Nernst observed.

\subsection{Potential of the SNB Model\label{snb}}

Schurtz, Nicola{\"{i}} and Busquet's (\capsformat{snb}) multigroup diffusion model for nonlocal electron heat transport \cite{SNB}
 has proven to be the most successful attempt to efficiently capture nonlocality in hydrodynamic simulations of \capsformat{icf} simulations.
This model calculates the contribution of  separate energy groups of electrons to the nonlocal heat flow by solving a set of independent inhomogeneous Helmholtz-like equations
\cite{SNB,BrodrickSNB}
and is able to capture both flux reduction and preheat effects 
but traditionally gives no prescription for nonlocal modifications to Nernst advection.
It has been implemented in a number of radiation-hydrodynamics codes used by national labs including
\capsformat{hydra} \cite{HYDRA}, \capsformat{chic} \cite{CHIC} and \capsformat{draco} \cite{Cao}.

By comparing the model's equation set to a simplified \capsformat{vfp} approach,
the authors were able to suggest a relation between the energy group contribution $H_g$ and 
the nonlocal perturbation to the isotropic part of the \capsformat{edf} $\delta f_0$.
If this relationship were accurate,
it would provide
a simple method of calculating corrections to the Nernst coefficient $\beta_\wedge$
by taking moments of the distribution function (see, for example, \cref{app:mom}).
However, we have recently shown that 
such a reconstruction
of the \capsformat{edf} from the \capsformat{snb} model does
not agree well with \capsformat{vfp} predictions \cite{SherlockSNB}. 
Particularly, the
\capsformat{snb} model appears not to account for the enhanced
return current predicted by \capsformat{vfp} codes in regions
where there is considerable preheat.
In retrospect, this is not surprising due to the approximate treatment of the electric field in the \capsformat{snb} model.
Therefore, it would be desirable to come up with a more reliable method of using the \capsformat{snb}
model to account for nonlocal Nernst advection.
One such approach is to use the observation that the ratio between the Nernst and heat flow velocities does not
depend greatly on nonlocality  \cite{HainesV2,KhoHainesPRL,*KhoHainesPoP}.

Specifically, if a good approximation for the nonlocal heat flow can be obtained 
 (such as from the \capsformat{snb} model) then we should be
able to estimate the nonlocal electric field by simply multiplying the former by the 
ratio expected in the local limit: $\beta_\wedge^{\mathrm{(Local)}}/e \kappa_\perp^{\mathrm{(Local)}} \equiv B \psi^{\mathrm{(Local)}}/P_{\mathrm{e}}$.
That is, $E_y^{\mathrm{(Nonlocal)}} \approx   (B \psi^{\mathrm{(Local)}}/P_{\mathrm{e}}) Q_x^{\mathrm{(Nonlocal)}}$.
The wide range of problems investigated here provide a perfect opportunity to thoroughly test 
whether this approximation is indeed accurate and reliable.

While a magnetised extension of the \capsformat{snb} model has been developed and is implemented in the \capsformat{chic} code \cite{SNBmag}, this has not yet been extensively tested against \capsformat{vfp} simulations.
We do not attempt to do this here. 
Instead, we simply apply the \textit{unmagnetised} model to simulations with low magnetisations ($\chi<0.03$) so that the heat flow and degree of nonlocality are not strongly affected by the presence of a magnetic field. 
The specific \capsformat{snb} implementation used here corresponds with the optimal one identified in \cite{BrodrickSNB}; this consists of
(i) imposing a scaling factor on the Krook electron-electron collision frequency of $r=2$, 
(ii) separating the electron-electron and electron-ion mfp's, and 
(iii) multiplying the electron-ion mfp by the collision fix $\xi$.

The bottom panels of \cref{fig:k2_0.1tesla,fig:z40_1tesla} present the results of using this approximation
for the helium 0.1 tesla and zirconium 1 tesla simulations.
Temperature and magnetic field profiles at 15 ps and 4 ps respectively were used to calculate the \capsformat{snb} heat flow before converting this to an estimate for $E_y$.
We observe that this method for obtaining the \capsformat{snb} electric field exhibits remarkable agreement with \capsformat{vfp}, closely matching both the degree of flux reduction and the preheat at very little additional computational cost.

At higher magnetisations, we were still able to test the claim that nonlocality does not affect the link between thermal conduction and Nernst advection by instead multiplying the \capsformat{vfp} heat flow by the  ratio $B \psi^{\mathrm{(Local)}}/P_{\mathrm{e}}$. 
This is depicted for the 2 tesla helium and the 10 tesla zirconium runs in the bottom panels of
\cref{fig:k2_2tesla,fig:z40_10tesla} respectively.
Again, the ratio method provides a good approximation for $E_y$, with the main discrepancy being a slight underestimate of the pre-Nernst on the right-hand side. This discrepancy arises due to the dependence of $\beta_\wedge$ on higher-velocity moments of the \capsformat{edf} than $\kappa_\perp$ (see \cref{app:mom}), making it more sensitive to nonlocal effects.

Our findings are summarised in \cref{fig:ratio_nonlocal} which presents the
nonlocal \capsformat{vfp} prediction for the dimensionless ratio $\psi^{\mathrm{(\capsformat{vfp}})} = P_\mathrm{e}E_y^{\mathrm{(\capsformat{vfp}})} / B Q_x^{\mathrm{(\capsformat{vfp}})} $ as a function of magnetisation for all temperature ramp relaxation simulations.
Profiles were extracted at 25 ps for the helium simulation and 5 ps for the zirconium.
It is shown that $\psi$ approximately follows the local prediction indicated by the dashed lines, 
clearly exhibiting a strong ionisation dependence that would not be captured by constant ratio approximations 
suggested by other authors \cite{HainesV2,Davies,LanciaNonlocalNernst}.
The prominent flick-ups seen at the low magnetisation end (left-hand side) of this figure
 correspond to increased reach of pre-Nernst as compared to pre-heat arising
from the dependence of $\beta_\wedge$ on higher-velocity moments of the \capsformat{edf}.

We also investigated the effectiveness of using a similar process to estimate the Righi-Leduc heat flow as $Q_y^{\mathrm{(Nonlocal)}} \approx (Q_x^{\mathrm{(Local)}}/Q_y^{\mathrm{(Local)}})Q_x^{\mathrm{(Nonlocal)}}$ in the middle panel of \cref{fig:k2_0.1tesla,fig:k2_2tesla,fig:z40_10tesla,fig:z40_1tesla}.
However, this approach 
underestimates the degree of flux-limitation  and
does not capture the high degree of preheat arising from 
the higher velocity moments used in calculating the Righi-Leduc heat flow.
Nevertheless, it is still a definite improvement on both the local Braginskii and flux-limiter approaches at lower magnetisations.

\section{Laser spot heating \label{sec:froula}}

As the degree of nonlocality for the temperature ramp relaxation problem
at high magnetisations
 was not sufficient to cause observable differences in the final temperature profiles, even after tens of collision times,
we also looked at a laser-heating  problem where the degree of nonlocality continually increases with time.
This included a
fully-ionised nitrogen plasma ($Z=7$) 
of
uniform electron density \SI[per-mode=reciprocal]{1.5e19}{\per\centi\meter\cubed}
and an assumed constant Coulomb logarithm of 7.5
being heated by 
a continuous \SI{6.3e13}{\watt\per\centi\meter\squared} laser
(no time envelope was applied in our treatment).
The intensity profile was  essentially uniform in the $y$ and $z$ directions and Gaussian in the $x$-direction with a full width at half maximum of \SI{150}{\micro\meter}.
Again ion motion was neglected.
This setup is based on an experiment performed by Froula \etal\ \cite{Froula}
that has previously been simulated with \capsformat{impact} by Ridgers \etal\ \cite{reemerge}.
Here we use the K2 code to correctly account for the effect of electron-electron collisions on the anisotropic part of the distribution function 
but restrict ourselves to a one-dimensional treatment for the sake of keeping runtimes short. (Thus, the beam profile is planar rather than cylindrical.)
Furthermore, the plasma had an initially uniform temperature of \SI{50}{\electronvolt}
which was slightly higher than the \SI{20}{\electronvolt} previously simulated by Ridgers \etal\ 
to reduce the number of velocity cells required.
A total of 250 velocity cells were used extending up to $v_\mathrm{max} = 25v_\mathrm{T}(\SI{50}{\electronvolt})$, corresponding to electrons with an energy of \SI{15.6}{\kilo\electronvolt}.
   The
spatial domain,
 consisting of 100 cells,
  extended to \SI{500}{\micro\meter} from the centre of the pulse
and again we used reflective boundary conditions.
Initially uniform magnetic fields of 4 tesla were imposed.
A timestep of \SI{2}{\femto\second} was used.

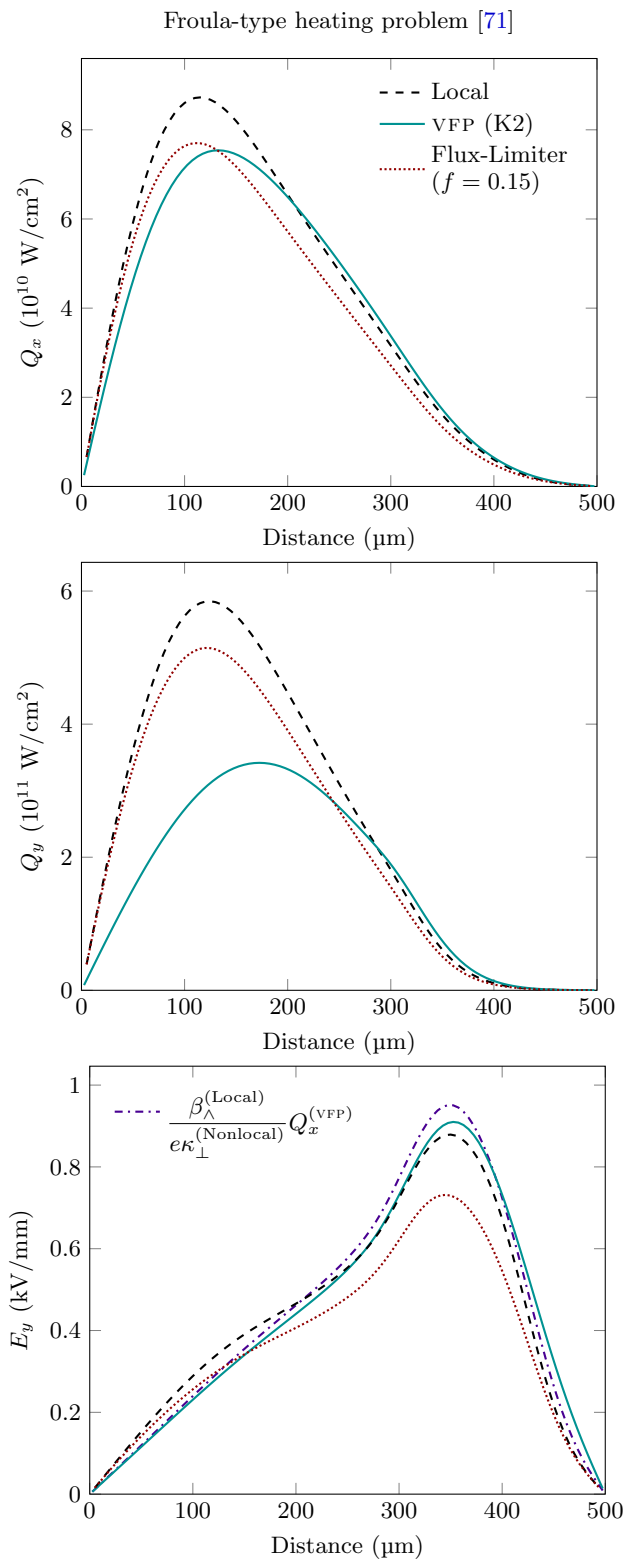
\begin{figure}
\begin{tikzpicture}
\begin{axis}[
title = {Froula-type heating problem \cite{Froula}},
xlabel = {Distance (\si{\micro\metre})},xmin=0.,xmax=500.,
ylabel = {$Q_x$ ($10^{10}$ \si{\watt\per\centi\metre\squared})},ymin=0.,
scaled y ticks={real:1e10},
ytick scale label code/.code={},
legend pos = north east
]
\addplot[local] table {data/Froula_4T_local_xmic_QxWcm2_600.0ps}; \addlegendentry{Local}
\addplot[vfp] table {data/Froula_4T_k2_xmic_QxWcm2_600.0ps}; \addlegendentry{\capsformat{vfp} (K2)}
\addplot[fluxlim] table {data/Froula_4T_fl.15_xmic_QxWcm2_600.0ps}; \addlegendentry{Flux-Limiter \\  ($f=0.15$)}
\end{axis}
\end{tikzpicture}

\begin{tikzpicture}
\begin{axis}[
xlabel = {Distance (\si{\micro\metre})},xmin=0.,xmax=500.,
ylabel = {$Q_y$ ($10^{11}$ \si{\watt\per\centi\metre\squared})},ymin=0.,
scaled y ticks={real:1e11},
ytick scale label code/.code={},
legend pos = north west,
]
\addplot[local] table {data/Froula_4T_local_xmic_QyWcm2_600.0ps}; 
\addplot[vfp] table {data/Froula_4T_k2_xmic_QyWcm2_600.0ps}; 
\addplot[fluxlim] table {data/Froula_4T_fluxlimperp_xmic_QyWcm2_600.0ps}; 
\end{axis}
\end{tikzpicture}

\begin{tikzpicture}
\begin{axis}[
xlabel = {Distance (\si{\micro\metre})},xmin=0.,xmax=500.,
ylabel = {$E_y$ (\si{\kilo\volt\per\milli\metre})},ymin=0.,
legend pos = north west,
]
\addplot[haines] table {data/Froula_4T_Hainesk2_xmic_EyMV_600.0ps};  \addlegendentry{$\displaystyle\frac{\beta_\wedge^{\mathrm{(Local)}}}{e\kappa_\perp^{\mathrm{(Nonlocal)}}}Q_x^{(\capsformat{vfp})}$}
\addplot[vfp] table {data/Froula_4T_k2_xmic_EyMV_600.0ps};
\addplot[local] table {data/Froula_4T_local_xmic_EyMV_600.0ps};
\addplot[fluxlim] table {data/Froula_4T_fl.15_xmic_EyMV_600.0ps};
\end{axis}
\end{tikzpicture}
\caption{\label{fig:froula}
Perpendicular and Righi-Leduc heat flows $Q_x$ (top), $Q_y$ (middle) and the Nernst-dominated out-of-plane electric field $E_y$ (bottom) 
after 600 ps K2 \capsformat{vfp} simulation for the Froula-type heating problem with an initial magnetic field of 4 tesla. 
Local and flux-limited profiles were postprocessed using the K2 temperature and magnetisation profiles. 
}
\end{figure}

Due to the initially uniform temperature profile, nonlocality did not begin to emerge until at least \SI{50}{\pico\second}.
This meant that the \capsformat{ctc} simulations could also be started from $t=0$. 
Despite nonlocality continually increasing, a flux-limiter of $0.15$ was found be a good match for the heat flow profile throughout most of the simulation.
However, even at the end of the \SI{600}{\pico\second} simulation the nonlocal reduction of the heat flow down the temperature gradient was only about 10\%. as shown in the top panel \cref{fig:froula}
While the electric field in the bottom panel experiences a similar reduction near the position of maximum heat flow, 
its peak is actually increased.
This may seem surprising but is explained by its occurrence in a region where preheat naturally occurs
(near the foot of the temperature gradient at \SI{400}{\micro\meter}),
thus enhancing the Nernst velocity due to a surplus of suprathermal electrons coming from the centre of the hot spot.

\begin{figure}
\begin{tikzpicture}
\begin{axis}[
xlabel = {Distance (\si{\micro\metre})},xmin=0.,xmax=500.,
ylabel = {$B_z$ (tesla)},ymin=0.,ymax=13.8,
legend pos = north west,
ytick distance=2,
]
\addplot[local] table {data/Froula_4T_nofluxlim_xmic_Btesla_600.0ps}; \addlegendentry{\capsformat{ctc} ($f_Q=\infty, f_\mathrm{N}=\infty$)}
\addplot[snb] table {data/Froula_4T_rf7.071_rN0_xmic_Btesla_600.0ps}; \addlegendentry{\capsformat{ctc} ($f_Q=0.15, f_\mathrm{N}=\infty$)}
\addplot[themePurple,dotted] table {data/Froula_4T_rf7.071_rN7.071_xmic_Btesla_600.0ps}; \addlegendentry{\capsformat{ctc} ($f_Q=0.15, f_\mathrm{N}=0.15$)}
\addplot[vfp] table {data/Froula_4T_k2_xmic_Btesla_600.0ps}; \addlegendentry{\capsformat{vfp} (K2)}
\end{axis}
\end{tikzpicture}

\begin{tikzpicture}
\begin{axis}[
xlabel = {Distance (\si{\micro\metre})},xmin=0.,xmax=500.,
ylabel = {$T_{\mathrm{e}}$ (keV)},ymin=0.,
]
\addplot[local] table {data/Froula_4T_nofluxlim_xmic_TkeV_600.0ps}; 
\addplot[snb] table {data/Froula_4T_rf7.071_rN0_xmic_TkeV_600.0ps}; 
\addplot[themePurple,dotted] table {data/Froula_4T_rf7.071_rN7.071_xmic_TkeV_600.0ps};
\addplot[vfp] table {data/Froula_4T_k2_xmic_TkeV_600.0ps}; 
\end{axis}
\end{tikzpicture}
\caption{\label{fig:bla1}Comparison of magnetic field and temperature profiles 
for the nitrogen heating problem
predicted by
K2 and 
 Classical Transport Code \capsformat{ctc} with different combinations of thermal and Nernst flux-limiters $f_Q, f_\mathrm{N}$ respectively.
  All profiles were evolved independently from an initial temperature of \SI{50}{\electronvolt}
  and magnetic field of \SI{4}{\tesla}
  for \SI{600}{\pico\second}.}
\end{figure}
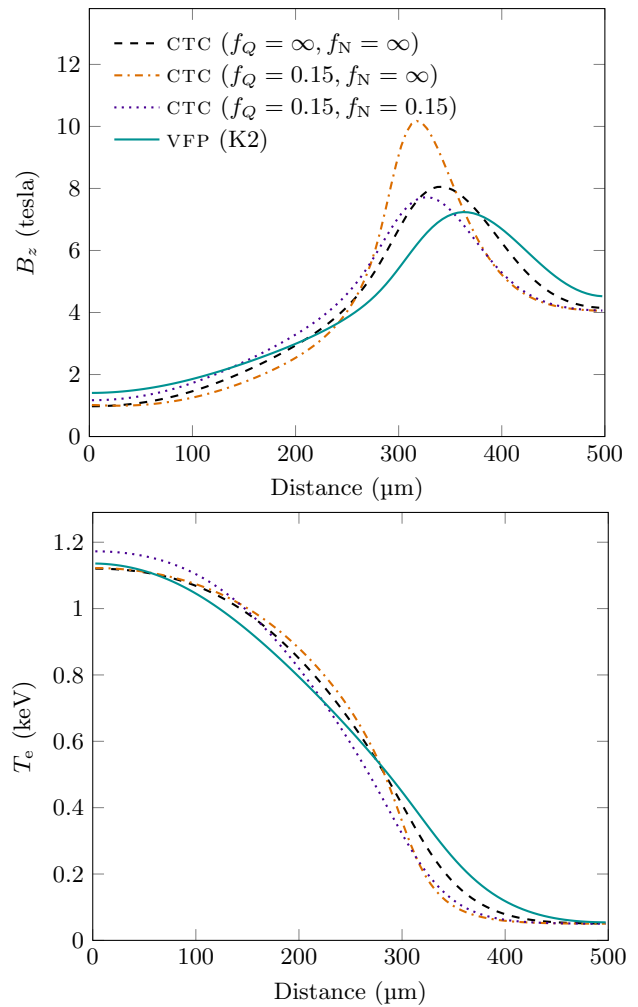

Comparing to the \capsformat{ctc} simulations we again find that applying only a thermal flux-limiter leads to an overamplification of the peak magnetic field (\cref{fig:bla1} top panel) at the end of the simulation by over 3 tesla (nearly 50\%).
In contrast, including a Nernst limiter reduces this error to less than 10\%.
However, we note that there is  still a nearly \SI{50}{\micro\meter} discrepancy in the location of the magnetic field crest due to the inability of the flux-limiter approach to incorporate the effect of pre-Nernst.
For this problem, there is a small but observable difference between the effect of the different approaches
in the final temperature profiles shown in the bottom panel of \cref{fig:bla1};
while inclusion of a Nernst limiter slightly increases the peak temperature
it  noticeably improves the prediction at 200--250 \si{\micro\metre}.
Again the Righi-Leduc heat flow was found to experience a much more severe flux-limitation (the peak flux was reduced by a factor of $\mathord{\sim}50\%$).

\section{Lineout from \capsformat{hydra} simulation with self-generated fields \label{sec:microdot}}

The effectiveness of linking Nernst advection to thermal conduction in a more realistic scenario
was confirmed by analysing a recent 
\capsformat{nif} viewfactor shot \cite{Viewfactor}
that employed a Mn/Co microdot \cite{Microdot} on the capsule surface for diagnostic purposes.
Radial lineouts were taken from a \SI{5}{\nano\second}  \capsformat{hydra} simulation 
that used a thermal flux-limiter of $f_Q = 0.15$ (see Fig 3 and the bottom panel of Fig 9 in \cite{Farmer2018}),
this employed the newly implemented \capsformat{mhd} suite (including Nernst) outlined in \cite{Farmer2017}. 
These lineouts were located 3 \si{\milli\meter} from the centre of the capsule,
starting in the low-density gas-fill at $r=0{}$
and ending just inside the partially heated hohlraum wall at $r={}$2.76 \si{\milli\meter},
and used to initialise a \SI{100}{\pico\second} \capsformat{vfp} relaxation simulation
using 1D planar geometry.
Again, only temperature and magnetic profiles were allowed to evolve
 while the density profile was fixed by neglecting ion hydrodynamics and using the zero current constraint.
In order to maintain consistency with the rest of this paper and to reinforce the fact that planar geometry was used
we will from this point on use the Cartesian coordinates $x, y,z$ in place of their cylindrical counterparts $r, z, (-)\phi$. The initial and  final ionisation, electron density, temperature and magnetic field profiles 
are illustrated in \cref{fig:MicrodotProfs}.

For this problem we used the \capsformat{impact} code to simplify treatment of the spatially-varying ionisation profile.
When calculating the local/flux-limited heat flow and Nernst profiles this enabled us to use the Lorentz limit ($Z=\infty$) transport coefficients with a multiplier of $\xi$ on all appearances of the collision time $\tau_{\mathrm{B}}$ instead of trying to interpolate between transport coefficients at other ionisations.
Note that there is a loss of accuracy incurred by making this simplification, particularly at low ionisations;
this error is  worst around $x={}$\SI{1}{\milli\metre}
where the ionisation is low and the magnetisation is not too high
leading to an underestimate of the Nernst velocity by a factor of approximately two (see \cref{fig:local_ratio}).
The  simulation setup included a spatial cell width of \SI{13.8}{\micro\meter} and a geometric velocity grid where the width in velocity-space of the highest energy cell (located at \SI{225}{\kilo\electronvolt}) was 30 times larger than the lowest energy cell.
We used a timestep of \SI{25}{\femto\second} and took the Coulomb logarithm to be constant at 4.1.

\begin{figure}
\begin{tikzpicture}
\begin{axis}[
    xlabel={$x$ (mm)},
   xmin=0.,xmax=2.8,
   ymin=-2.,ymax=125.,
   legend pos = north west,legend columns=2
   ]
\addplot[themeGray,densely dotted,forget plot] table [y expr=\thisrowno{1}*10] {data/microdot_entire_impact_x_mm_T_keV_0ps};
\addplot[fluxlim] table [y expr=\thisrowno{1}*10] {data/microdot_entire_impact_x_mm_T_keV_100.0ps};
\addlegendentry{$T_{\mathrm{e}}$ (100 eV)};
\addplot[themePurple,dashdotted] table {data/microdot_entire_impact_x_mm_ne_cmm3m20};
\addlegendentry{$n_{\textsf{e}}$ ($10^{20}$ cm\textsuperscript{-3})};
\addplot[local] table {data/microdot_entire_impact_x_mm_Z};
\addlegendentry{$Z$};
\addplot[themeGray,forget plot] table {data/microdot_entire_impact_x_mm_Bz_tesla_0ps}; 
\addplot[vfp] table {data/microdot_entire_impact_x_mm_Bz_tesla_100.0ps};
\addlegendentry{$B_z$ (tesla)}
\end{axis}
\end{tikzpicture}
\begin{tikzpicture}
\begin{axis}[
    xlabel={$x$ (mm)},
    ylabel={Magnetisation ($\chi$)},
   xmin=0.,xmax=2.8,
   ymin=0.,ymax=4.8,
   ]
\addplot[black]  table {data/microdot_entire_impact_x_mm_chi_100.0ps}; 
\end{axis}
\end{tikzpicture}
\caption{\label{fig:MicrodotProfs}
Spatial profiles of plasma temperature, electron density, ionisation, magnetic field, (top) and magnetisation (bottom) profiles 
based on a lineout after  from a \SI{5}{\nano\second} \capsformat{hydra}${}+{}$\SI{100}{\pico\second} \capsformat{impact} simulation. The initial temperature and magnetic field profile input to \capsformat{impact} are shown in grey.}
\end{figure}
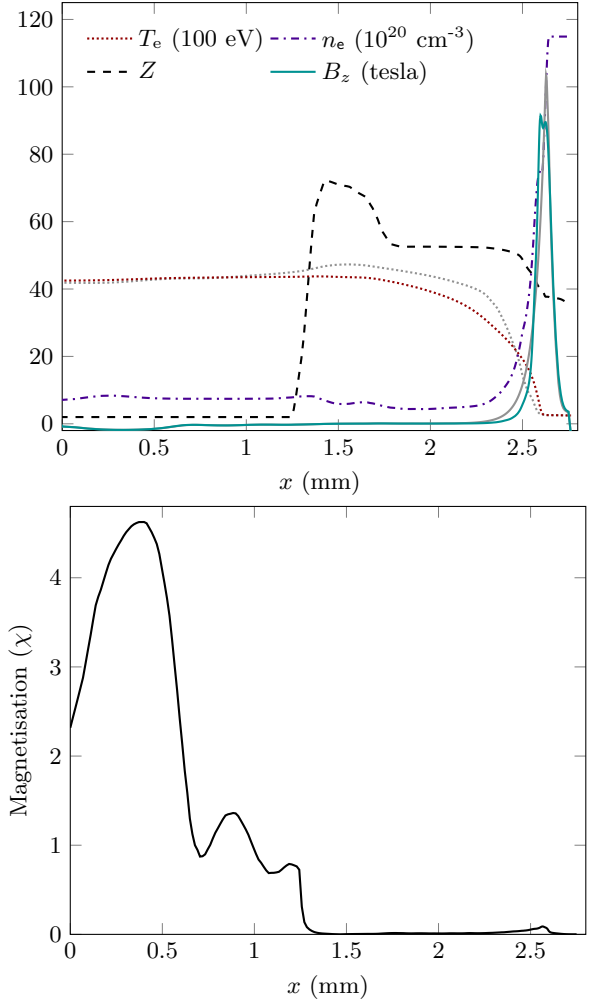
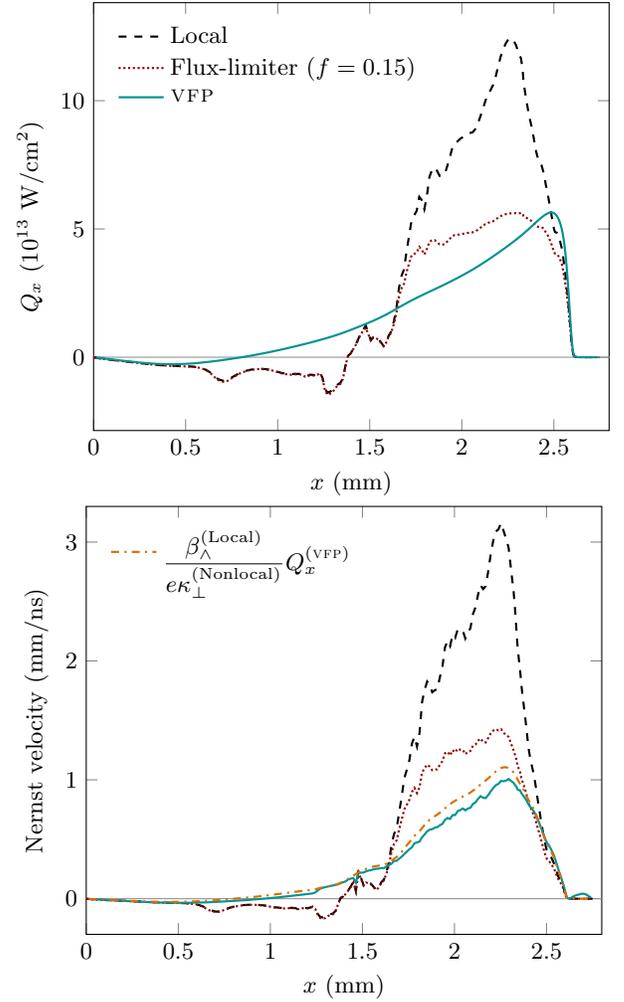
\begin{figure}
\begin{tikzpicture}
\begin{axis}[
    xlabel={$x$ (mm)},
   ylabel={$Q_x$ ($10^{13}$ \si{\watt\per\centi\meter\squared})},
   xmin=0.,xmax=2.8,
scaled y ticks={real:1e13},
ytick scale label code/.code={},
legend pos = north west]
\addplot[themeGray,forget plot,thin] {0.0};
\addplot[local] table {data/microdot_entire_impact_x_mm_Qxloc_Wcm2_100.0ps};
\addlegendentry{Local}
\addplot[fluxlim] table {data/microdot_entire_impact_x_mm_Qxfl.15_Wcm2_100.0ps};
\addlegendentry{Flux-limiter ($f=0.15$)}
\addplot[vfp] table {data/microdot_entire_impact_x_mm_Qx_Wcm2_100.0ps};
\addlegendentry{\capsformat{vfp}}
\end{axis}
\end{tikzpicture}
\begin{tikzpicture}
\begin{axis}[
    xlabel={$x$ (mm)},
   xmin=0.,xmax=2.8,
   ylabel={Nernst velocity (mm/ns)},
ymin=-.3,ymax=3.3,
   xmin=0.,xmax=2.8,
legend pos = north west]
\addplot[themeGray,forget plot,thin] {0.0};
\addplot[local, forget plot] table {data/microdot_entire_impact_x_mm_vNloc_mmperns_100.0ps};
\addplot[fluxlim, forget plot] table {data/microdot_entire_impact_x_mm_vNfl.15_mmperns_100.0ps};
\addplot[vfp,forget plot] table {data/microdot_entire_impact_x_mm_vN_mmperns_100.0ps};
\addplot[themeOrange,dashdotted] table {data/microdot_entire_impact_x_mm_vNHaines_mmperns_100.0ps};
\addlegendentry{$\displaystyle\frac{\beta_\wedge^{\mathrm{(Local)}}}{e\kappa_\perp^{\mathrm{(Nonlocal)}}}Q_x^{(\capsformat{vfp})}$}
\end{axis}
\end{tikzpicture}
\caption{\label{fig:MicrodotvN} The heat flow (top) and Nernst velocity (bottom) for the \capsformat{hydra} lineout after \SI{100}{\pico\second} \capsformat{impact} simulation.}
\end{figure}

The magnitude of the magnetisation
 is illustrated in the bottom panel of \cref{fig:MicrodotProfs}, but note that this conceals the reversal of the magnetic field  at about \SI{1.45}{\milli\meter}.
It is shown that,
 despite the magnetic field reaching megagauss levels in the hohlraum wall,
 the degree of magnetisation is actually quite low due to the very high collisionality in this region.
Conversely, the highest levels of magnetisation (exceeding unity) are reached near the centre ($x=0$) of the lineout, deep in the hot gas-fill. 
Therefore, instead of plotting the axial electric field $E_y$, which increases almost linearly with magnetic field and would be largest in the hohlraum wall where magnetisation effects are unimportant,
we instead consider  the Nernst velocity $v_\mathrm{N}=E_y/B$ itself in the bottom panel of \cref{fig:MicrodotvN}.
Note that the magnetisation profile does not change noticeably over the \SI{100}{\pico\second} simulation as
it is highest in a region of relatively homogeneous temperature.

Reduction of the Nernst velocity relative to the local prediction between $x={}$\SI{1.6}{\milli\meter}  and \SI{2.5}{\milli\meter} shows that magnetic field is advected into the hohlraum wall at a slower rate than expected,
reducing the amplification of the magnetic field in a similar manner to the previous test problems.
Relocalisation due to the high magnetic field means that there is a very low degree of preheat into the hohlraum beyond $x={}$\SI{2.5}{\milli\metre}.
Closer to the centre
we see a reversal of the Nernst velocity compared to the local prediction, 
meaning that the magnetic field is allowed to climb up the temperature gradient.
This is again another effect that could not be captured by a flux-limiter (red dotted).
Here a Nernst flux-limiter of 0.15 (as calculated by postprocessing the \SI{100}{\pico\second} profiles with \capsformat{ctc})
seems slightly conservative, and a lower value would be necessary to capture the high degree of flux reduction between \SI{1.5}{\milli\meter} and \SI{2}{\milli\meter},
but is nevertheless an improvement on the pure Braginskii approach.
Using the new method
of multiplying the local Nernst term by the ratio between the nonlocal \capsformat{vfp}  and local Braginskii heat 
flows
is highly accurate within a radius of approximately \SI{2.61}{\milli\meter}, 
at which point resistive diffusion becomes more important.
For the case of the heat flow shown in the \textit{top} panel of \cref{fig:MicrodotvN} a flux-limiter of 0.15 gets the peak about right, but again misses the nonlocal flux reversal observed and overestimates the heat flow near to $x={}$\SI{2}{\milli\meter}.

\section{Discussion}

The findings in this paper confirm, generalise and extend a number of previous observations
\cite{LanciaNonlocalNernst,HainesCanJPhys,HainesV2,Archis,read,Davies}
 about the effect of nonlocality on Nernst advection.
 Nonlocal limitation of the Nernst velocity reduces both the rate at which the magnetic field cavitates from hot regions of the plasma and the associated convective amplification of the magnetic field at the foot of the temperature gradient.
 It is the latter effect that is especially affected by nonlocality due to the additional effect of suprathermal electrons allowing the magnetic field to spread out further than would be expected from a local prediction;
 a phenomenon that could never be replicated by a flux-limiter approach.

By studying a wide range of problems and ionisations, we fully confirm the claim made by Haines \cite{HainesV2} that the relationship between thermal conduction and Nernst advection should not be greatly affected by nonlocality.
This allows for a simple method of using the prediction from a nonlocal heat flow model (such as the \capsformat{snb}) to calculate the Nernst velocity $v_{\mathrm{N}}\approx \psi^{(\mathrm{(Local)})}Q_x/P_{\mathrm{e}}$,
where $\psi^{\mathrm{(Local)}}= P_{\mathrm{e}}\beta_\wedge^{\mathrm{(Local)}}/eB\kappa_\perp^{\mathrm{(Local)}}$ is calculated using the Epperlein and Haines coefficients \cite{EppHaines}.
Crucially, this differs from previous suggestions that treat $\psi$ as a constant (either 2/5 \cite{Davies,Archis,reemerge,HainesV2} or 2/3 \cite{LanciaNonlocalNernst}) potentially resulting in errors of up to 80\% (see \cref{fig:local_ratio}).

Our  analysis  on the effects of using different combinations of flux-limiters,
suggests that if a more sophisticated approach is not available
then
it is safest to use identical flux-limiters on heat flow and Nernst advection
thus avoiding the introduction of an additional tunable parameter.
Specifically, this should be applied in such a way that relative reductions in $Q_x$ and $E_y$ are equivalent.

While it may seem comforting that nonlocal modifications to the Nernst velocity
do not seem to have significant knock-on effects
(such as on the evolution of temperature profiles) in the problems studied,
this may not be universally true.
Firstly, the keV-scale reductions in plasma temperature associated with including Nernst advection in indirect-drive \capsformat{hydra} simulations observed by Farmer \etal\ \cite{Farmer2017}
suggest that limiting Nernst should increase the final temperature by a non-negligible amount if applied throughout the entire simulation (as opposed to the rather limited \SI{100}{\pico\second} considered in \cref{sec:microdot}).
These increases in the plasma temperature due to Nernst limitation could reduce the absorption of laser energy due to inverse bremsstrahlung,
if such temperature rises were concentrated in the gold bubble the resulting ability of the inner beams to deposit their energy nearer the hohlraum midplane could lead to a more prolate implosion \etal\ \cite{Farmer2017}.
However, 
the \capsformat{mhd} simulations performed by Farmer \etal\ \cite{Farmer2017} 
with the Nernst term disabled essentially put a bound on the degree to which 
Nernst limitation
could affect the x-ray drive,
meaning that nonlocal effects are unlikely to fully explain the drive deficit.
Nevertheless, 
nonlocality of Nernst advection could be more important for experiments involving 
externally imposed fields, as was the case for the direct-drive shot studied by
Davies \etal\ \cite{Davies}
where  modifying the Nernst limiter 
led to discernible differences in the neutron yield and ion temperature.
Finally, the reversal of Nernst advection observed in \cref{sec:microdot}
may have unexpected effects such as pinning the magnetic field to the hohlraum wall,
 and somewhat reducing the thermal insulation in the interior of the corona.

One omission that was made in all \capsformat{vfp} simulations presented in this paper was the neglection of ion hydrodynamics.
This was to simplify the analysis by focussing only on heat and magnetic field transport.
Such an assumption is unlikely to greatly affect the resulting physics over the timescales studied here.
For example, rerunning the flux-limited \capsformat{ctc} simulations with ion motion included for the helium temperature ramp relaxation problem revealed that the resulting change in the electron density over \SI{300}{\pico\second} would not exceed 5\%.
While this has slight knock-on effects for the evolution of the magnetic field,
decreasing the degree of amplification and cavitation by up to 5\%,
the consequence for the temperature profile is negligible.





It is worth pausing to consider the potential importance of nonlocal effects
on other transport phenomena in the magnetised regime.
Perhaps the strongest candidate for further investigation
is the Righi-Leduc heat flow due to its dependence on very high velocity moments of the \capsformat{edf} (e.g.\ $\langle V^{12} \rangle$, as elucidated in \cref{app:mom}). 
Severe flux-limitation of the Righi-Leduc heat flow, as  observed here, could potentially alleviate some of the hot spot cooling recently observed in simulations by Walsh \etal\ in the 
stagnation phase of indirect-drive implosions \cite{WalshPRL}
(although the degree of nonlocality in their simulations may not have been sufficiently high enough for a significant alleviation).
Also, the field compressing magneto-thermal instability involves the coupling of Righi-Leduc heat flow with Nernst advection \cite{CTC}
and the work here could help achieve a better understanding of how it behaves under non-local conditions without performing expensive \capsformat{vfp} calculations.
However,
the absence of an obvious link with the perpendicular heat flux
means that there is no simple way of accounting for nonlocal effects on the Righi-Leduc heat flow 
without having to resort to the addition of a new \textit{independent} flux-limiter
or a more sophisticated reduced nonlocal model capable with stronger links to the \capsformat{edf} itself
 (such as as the M1 model \cite{Dario,Dariomag} including B-fields; whose accuracy has yet to be fully established).
 
 Less affected by nonlocality is the usually negligible effect of resistive diffusion which relaxes steep magnetic field gradients.
This is due to the relevant transport coefficient $\alpha_\perp$ only depending on the fifth velocity moment $\langle V^5\rangle$ of the distribution function \cite{Epperlein94} (see \cref{app:mom}).
 
 One phenomena not investigated here is the the self-generation of magnetic fields by the Biermann battery effect that occurs in presence of transverse density and temperature gradients.
 And Kingham and Bell \cite{NonlocalBiermann} have shown that nonlocality can lead to analogous magnetic field generation even in the complete absence of density gradients.
 Further work is therefore required to consider the importance of and develop models for these nonlocally generated fields.

%

\section{Conclusions}
In this paper we find that the 
advection of magnetic fields down steep temperature gradients due to the Nernst effect
 experiences both a nonlocal flux reduction 
as well as a significant degree of
`pre-Nernst', which transports magnetic field beyond the temperature gradient.
Our simulations show both these effects working together to reduce the build-up of magnetic field and smearing it out into colder regions.
If these effects are not taken into account it is possible that overamplification of the magnetic field could lead to unphysical thermal transport barriers.
A simple but effective method of obtaining a reliable nonlocal prediction for the Nernst thermoelectric coefficient from a nonlocal heat flow model,
one that does not require developing a new highly sophisticated model capable of accurately approximating the entire \capsformat{edf},
 is $\beta_\wedge^{\mathrm{(Nonlocal)}}=\kappa_\perp^{\mathrm{(Nonlocal)}}\beta_\wedge^{\mathrm{(Local)}}/\kappa_\perp^{\mathrm{(Local)}}$.
\appendix

\section{Integral form for transport coefficients\label{app:mom}}
In the low magnetisation $\chi\rightarrow 0$ and Lorentz ($Z=\infty$) limits the integral form given by Epperlein \cite{Epperlein94} for the normalised
transport coefficients discussed in the paper take the following form:
{
\allowdisplaybreaks
\begin{align}
\kappa_\perp^c &= \frac{8\sqrt{\pi}}{9}\Big(\langle V^9 \rangle - \frac{\langle V^7\rangle^2}{ \langle V^5\rangle } \Big), \\
\kappa_\wedge^c &= \frac{8\sqrt{\pi}}{9}\Omega\Big(\langle V^{12} \rangle - 2\frac{\langle V^{10}\rangle \langle V^7\rangle}{\langle V^5\rangle}-\frac{\langle V^7\rangle^2\langle V^8\rangle}{ \langle V^5\rangle^2}\Big) , \\
\beta_\wedge^c &= \Omega\Big(\frac{\langle V^{10}\rangle}{\langle V^5\rangle}-\frac{\langle V^8\rangle\langle V^7\rangle}{\langle V^5\rangle^2}\Big),
\end{align}
}
\noindent where Epperlein's notation $\langle V^n \rangle=\int_0^\infty \sqrt{2} \pi v_{\mathrm{T}}^3 v^n f_0(v) / n_{\mathrm{e}}\diff v$ denotes moments of the isotropic part of the distribution function $f_0$  and $\Omega = 4\chi/3\sqrt{\pi}$. 

\section*{Acknowledgements}
We would like to acknowledge the assistance of J Fuchs and A Grissoulet in understanding their work with L Lancia \cite{LanciaNonlocalNernst} and R Riquier \cite{riqui2016}.

All data used to produce the figures in this work,
along with other selected supporting data, can be
found at \href{http://dx.doi.org/10.15124/58594ac4-8752-498c-8721-d5839082ed49}{dx.doi.org/10.15124/58594ac4-8752-498c-8721-d5839082ed49}.

J P Brodrick would like to thank the IOP RSCF for part-funding his attendance at the Anomalous Absorption meeting where this work was originally presented.

This work is funded by EPSRC grants EP/K504178/1 and EP/M011372/1.
This work has been carried out within the framework of the EUROfusion Consortium and has received funding from the Euratom research and training programme 2014--2018 under grant agreement No 633053 (project reference CfP-AWP17-IFE-CCFE-01). The views and opinions expressed herein do not necessarily reflect those of the European Commission. 
This work was performed under the auspices of the U.S. Department of Energy by Lawrence Livermore National Laboratory under Contract DE-AC52-07NA27344. 
This work was supported by Royal Society award IE140365. 
Partial support was provided by NSF under Grant No. ACI-1339893 and by DOE under Grant No. DE-NA0002953.

\bibliography{reference}{}

\end{document}